# Benzo-bis(imidazole) Self-Assembled Monolayers Molecular Junctions in *meta* or *para* Conformation: Effects of Protonation on the Electrical and Thermal Conductances.

Sergio Gonzalez-Casal,[1,§] Simon Pascal,[2,3] Olivier Siri[2] and Dominique Vuillaume.[1*]

*1) Institute for Electronics Microelectronics and Nanotechnology (IEMN), CNRS, Villeneuve d'Ascq, France.*
*2) Centre Interdisciplinaire de Nanoscience de Marseille (CINaM), CNRS, Luminy, France.*
*3) Chimie et Interdisciplinarité, Synthèse, Analyse, Modélisation( CEISAM), CNRS, Nantes, France.*

*§ Now at Institut Néel, Université Grenoble Alpes, Grenoble, France.*

** Corresponding author: dominique.vuillaume@iemn.fr*

**Abstract.**

We report the thermal conductances of molecular junctions made of self-assembled monolayers of benzo-bis(imidazole) molecules, without side groups or functionalized with two phenylamine side groups. In the two cases, when the molecules are connected to the electrodes by thiol anchoring groups in the *meta*-position, the thermal conductance is decreased compared to the same molecules connected in the *para*-position (≈ 16-29 nW/K and ≈ 37-40 nW/K, respectively) in agreement with the theoretically predicted phonon interference effect in molecular junctions. Upon protonation, the thermal conductances of the *meta*-connected molecular junction increase by about 50% (reversible behavior upon deprotonation). We propose that this behavior is related to modifications of the structural organization of the molecules in the monolayer, which modifies the thermal conductance at the molecule/electrode interfaces. The electrical conductance is lower for the *meta*-connected molecule than for the *para*-connected one, due to destructive quantum interferences, as expected and reported for other molecular junctions. We report that the electrical conductance further decreases (reversibly) upon protonation. The energy position of the molecular orbital involved in the electron transport is not modified by the protonation and the decrease in current is related to changes in the molecule organization in the monolayer, which modulate the electronic coupling energy at the molecule/electrode interfaces.

## 1. INTRODUCTION.

It is known that the electron transport properties of molecular junctions (MJs) are different when the molecules are connected to the electrodes in the *para* or *meta* positions. Due to the presence of destructive quantum interferences (DQI) in the MJs, the *meta*-connected MJs have lower electrical conductances than the *para*-connected MJs.[1-11] Most of the time, the *para*/*meta* effect has been studied for "simple" or "model" molecules (e.g., OPE, OPV, oligobenzene…). However, several studies have demonstrated significant electrical conductance dynamic switching due to the conformational control of QI in the MJs (*i.e.*, switching between DQI and constructive QI) by external stimuli: electrical gating,[12] light irradiation,[13, 14] electrochemical or chemical gating,[15-18] mechanical strain.[19, 20] The effect of DQI (whatever their origin) on the thermoelectric behaviors of MJs is generally to increase the Seebeck coefficient.[7, 21-27] In contrast, studies on the effects of *para* vs. *meta* connectivity on thermal conductance are scarce. Several theoretical studies predicted a reduction of the thermal conductance for a *meta*-connected molecule in the MJs or *meta*-coupling inside the molecule,[28-31] albeit the behavior might be inverted (thermal conductance of meta-connected MJs larger than for the *para* connectivity) depending on the geometrical detail of the S-Au contact (atop vs. hollow Au contact).[32] However, the experimental proof has remained elusive until very recently. In a SThM single-molecule experiment, Yelishala et al. demonstrated a reduction by a factor *ca.* 2 of the thermal conductance for a *meta*-OPE3 MJ (≈ 17 pW/K) versus the *para*-OPE3 (≈ 28 pW/K).[33]

Herein, we study the *para*/*meta* connectivity effect on the electrical and thermoelectric properties of switchable (pH stimulus) MJs. Benzo-bis(imidazole) molecules with thiol anchoring groups in the *para* and *meta* positions were synthesized. Two series of molecules were considered without side groups or with phenylamine side groups substituted on the bis(imidazole) backbone. The molecules were equipped with (trimethylsilyl)ethane protected thiol anchoring groups in *para* and *meta* positions, leading to four compounds: **1-*para***, **1-*meta***, **2-*para*** and **2-*meta*** (Fig. 1).

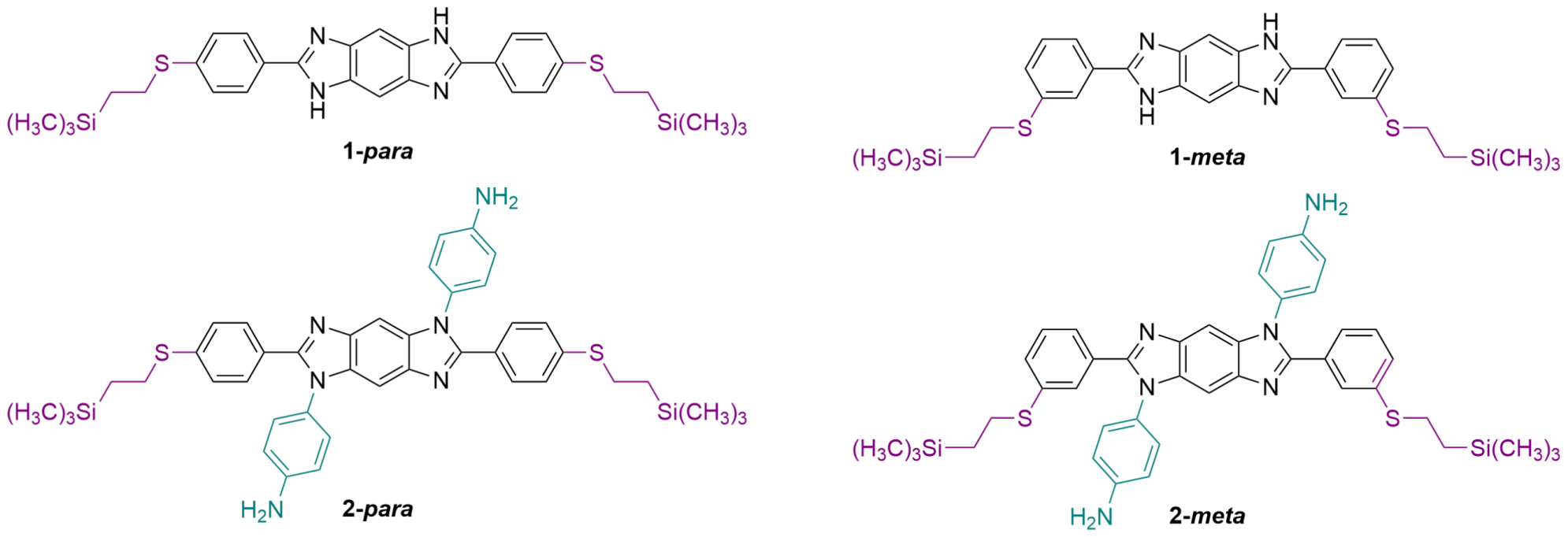

***Figure 1**. Structures of the four benzo-bis(imidazole) derivatives.*

Self-assembled monolayers (SAMs) from the four molecules were formed on ultra-flat template-stripped $^{TS}$Au electrodes and measured by C-AFM (electrical conductance) and SThM (thermal conductance). We compared two series of benzo-bis(imidazole) and we found, in the two cases, that the electron and thermal conductances of the SAMs are lower for the *meta*-connected MJs (SAM conductances ≈ $10^{-9}$-$4.1\times10^{-9}$ S for *meta*-MJs and ≈ $4.2\times10^{-9}$-$2.4\times10^{-8}$ S for *para* conformation, SAM thermal conductances ≈ 37-40 nW/K and ≈ 16-29 nW/K for *para* and *meta* connectivity, respectively). Upon protonation/deprotonation, the electronic conductances of the *meta*-connected MJs systematically decreased (by a factor ≈ 2-15), while in the *para* conformation,[34] the behavior depends on the nature of the side groups (the conductance increases for **1-*para*** and it decreases for **2-*para*** MJS). The SAM thermal conductances of the *meta*-connected MJs increased upon protonation (by about 50%). All these switching behaviors are reversible upon deprotonation. The fact that the thermal conductance of the *meta*-connected SAMs is affected by the protonation/deprotonation is tentatively related to the modification of the structural organization of the molecules in the *meta*-connected SAMs, which modifies the thermal conductance at the SAM/C-AFM tip interface.

## 2. EXPERIMENTAL.

### ***2.1. Synthesis of molecules***.

The synthesis of **1-*para*** and **2-*para*** has been already reported, as well as the effect of protonation/deprotonation on the electrical properties of the MJs based on these molecules.[34] Here, we have developed the one-step synthesis of the molecules **1-*meta*** and **2-*meta*** from aldehyde **2** (see in the Supplementary information) by condensation with 2,5-diamino-1,4-benzoquinonediimine (**DABQI**) and Brandowski's base (**BB**), respectively (Scheme 1). The target molecules **1-*meta*** and **2-*meta*** are obtained as a beige solid with 72% and 59% yields, respectively. These compounds have been fully characterized by NMR, FTIR and HRMS (supplementary information, Figs. S1-S16).

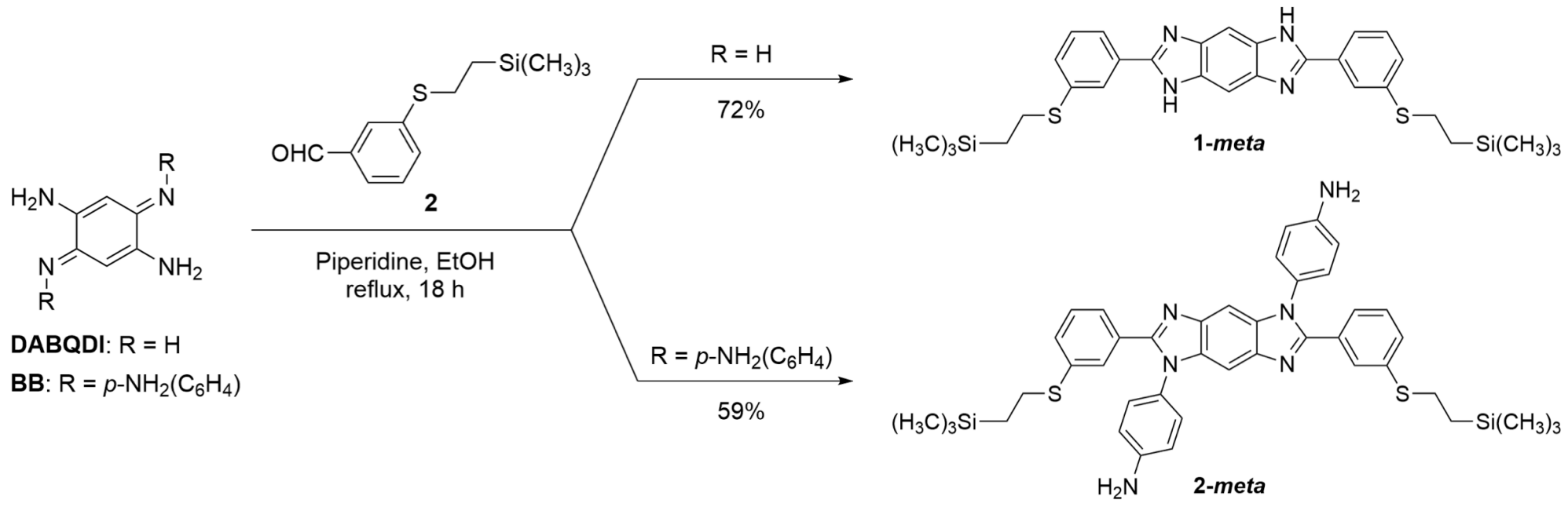

***Scheme 1**. Synthesis of **1-meta** and **2-meta** benzo-bis(imidazole) derivatives.*

## 2.2. *Fabrication of molecular junctions*.

The MJs were fabricated by forming a self-assembled monolayer (SAM) on an ultra-flat template-stripped gold substrate ($^{TS}$Au). The $^{TS}$Au substrates were prepared according to already reported methods.[35-37] The root mean square (rms) roughness of these substrates is ≈ 0.3-0.4 nm as measured by atomic force microscopy (AFM) in our previous works.[38] In a glove box ($N_2$ filled), the freshly prepared $^{TS}$Au substrates were dipped into a 1 mM solution of the molecules in DMSO:ethanol (50:50) for 3 days (in the dark) with $NBu_4F$ added to unprotect the sulfur anchor. The samples were copiously rinsed with deionized water and ethanol. The thicknesses of the SAMs were characterized by spectroscopic ellipsometry (see details in the supplementary information).

## 2.3. *Current-voltage (I-V) characterization by conductive-AFM*.

We measured the current-voltage (I-V) characteristics of the devices by conductive-AFM (C-AFM) with a Dimension Icon microscope (Bruker) equipped with a conductive PtIr metal-plated tip (model RMN-12PT400B from Bruker). The C-AFM is installed in an air-conditioned laboratory ($T_{amb}$ = 22.5 °C, a relative humidity of 35-40%). We applied a low loading force (3-5 nN) and 200 I-V traces were recorded on a square grid of 10 × 10 points (pitches of 50 to 100 nm). A back and forth I-V curve was acquired at each point. These measurements were repeated at several locations on the SAMs, leading to a raw dataset of several hundreds of I-V traces. The raw datasets were inspected and unreliable I-V traces were discarded from the analysis (see details in the supplementary information). The final number of I-V traces is indicated in the figures.

***2.4. Thermal conductance by scanning thermal microscope (SThM)***.

SThM[39, 40] were carried out with a Bruker ICON instrument equipped with the Anasys SThM module and in an air-conditioned laboratory (22.5°C and a relative humidity of 35-40 %). We used Kelvin NanoTechnology (KNT) SThM probes with a Pd thin film resistor in the probe tip as the heating element (VITA-DM-GLA-1). The SThM tip is inserted into a Wheatstone bridge; the heat flux through the tip is controlled by the DC voltage applied on the Wheatstone bridge ($V_{WB}$, typically 0.6-1.1 V). The tip temperature, $T_{tip}$, is obtained by measuring the output voltage of the Wheatstone bridge, knowing the transfer function of the bridge, the gain of the voltage amplifier and the calibrated linear relationship between the tip resistance and the tip temperature. The null-point SThM (NP-SThM)[41] was used at several locations on the SAMs. This differential method is suitable to remove the parasitic contributions (thermal radiation, air conduction and convection)[40, 42]: at the contact (C) both the sample and parasitic thermal contributions govern the tip temperature, whereas, just before physical tip contact (NC), only the parasitic thermal contributions are involved (more details and the calibration protocol are given in the supplementary information). The temperature difference $T_{NC}$ - $T_C$ is a measure of the sample thermal conductivity.[41] We have verified in previous work the robustness of the SThM "null-point" method and of our experimental setup by comparison with other standard methods (photothermal radiometry, flash laser analysis, and micro-Raman), which are not sensitive to the tip-sample interface issues. There was good agreement among all these techniques (relative errors ≲ 10 %).[43]

***2.5. Protonation and deprotonation of the MJs***.

For the protonation the SAMs are briefly (few tends of seconds) exposed to vapors of a diluted HCl solution and briefly exposed to $NEt_3$ vapors for deprotonation, as done in our previous work.[34] The samples are held at 2–3 cm above the surface of the solution.

## 3. RESULTS.

Note that **1-para** and **2-*para*** MJs were previously studied in detail,[34] except their thermal conductances, which are reported here. Other properties for these MJs are taken from ref 34.

***3.1. Structure properties of the SAMs***.

We have observed a reversible change in the SAM thickness upon protonation/deprotonation from 0.6 nm (pristine) to 1.1 nm (after protonation) and back to 0.6 nm after deprotonation (both for **1-*meta*** and **2-*meta***). The molecules are initially largely tilted (≈ 66° from the surface normal; the geometric optimized length of the molecules is ≈ 1.5-1.6 nm, S to S atoms, see details in the supplementary information, Fig. S17) and straighten up (tilt angle ≈ 45°) upon protonation. Note that the measured thicknesses of **1-*para*** and **2-*para*** did not change upon protonation/ deprotonation (1.8-1.9 nm for **1-*para*** and 1.3-1.5 nm for **2-*para***.[34] The geometric optimized length of the molecules is ≈ 1.8 nm, S to S atoms, Fig. S17. Clearly, the *meta*-connected molecules in the SAMs are less densely packed, since the SAM thicknesses are thinner than for the *para*-connected molecules, albeit the lengths of the molecules are almost similar (see the supplementary information, data in Table S1).

***3.2. Electrical conductance (C-AFM)***.

Figures 2 and 3 shows the I-V datasets and the current distribution histograms (at ± 0.2 V) of the **1-*meta*** and **2-*meta*** MJs (respectively) in the pristine state, after protonation and deprotonation. Figure 4-a summarizes the evolution of the current and compares with the data for the **1-*para*** and **2-*para*** MJs already published.[34] In the pristine state, the electrical conductance of **2-*para*** (and **2-*meta***) is larger than that of **1-*para*** (and **1-*meta***) by a factor ≈ 5, respectively (Fig. 4). The MJs with the *meta* conformation have a lower electron conductance (as usually reported).[1-3, 5, 6, 8, 11, 44] After protonation, the conductances of **1-*meta*** and **2-*meta*** decreased (slightly for **1-*meta*** by a factor ≈ 2, more significantly for **2-*meta*** by a factor ≈ 15), while we previously reported that the MJ conductance increased (decreased) for **1-*para*** (**2-*para***), respectively.[34] These conductance switching are reversible upon deprotonation.

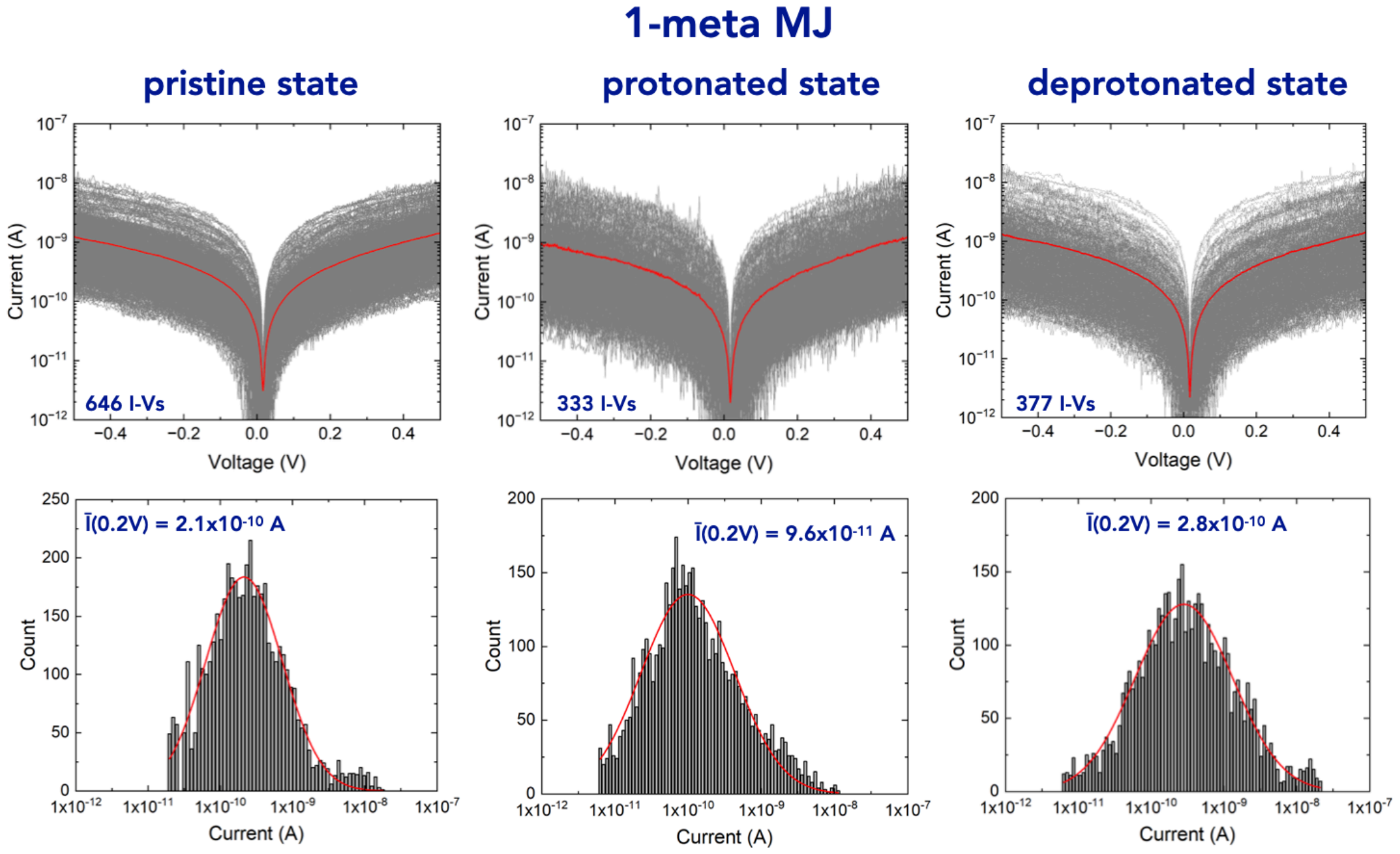

***Figure 2***. *I-V datasets and corresponding histograms of the currents (at 0.2 V) for the* ***1-meta*** *MJs: pristine state, after protonation and deprotonation. In the datasets, the red lines are the mean Ī-V curves. In the current histograms, the red lines are the fits by a log-normal distribution. The mean values of the current, Ī, are indicated in the panels.*

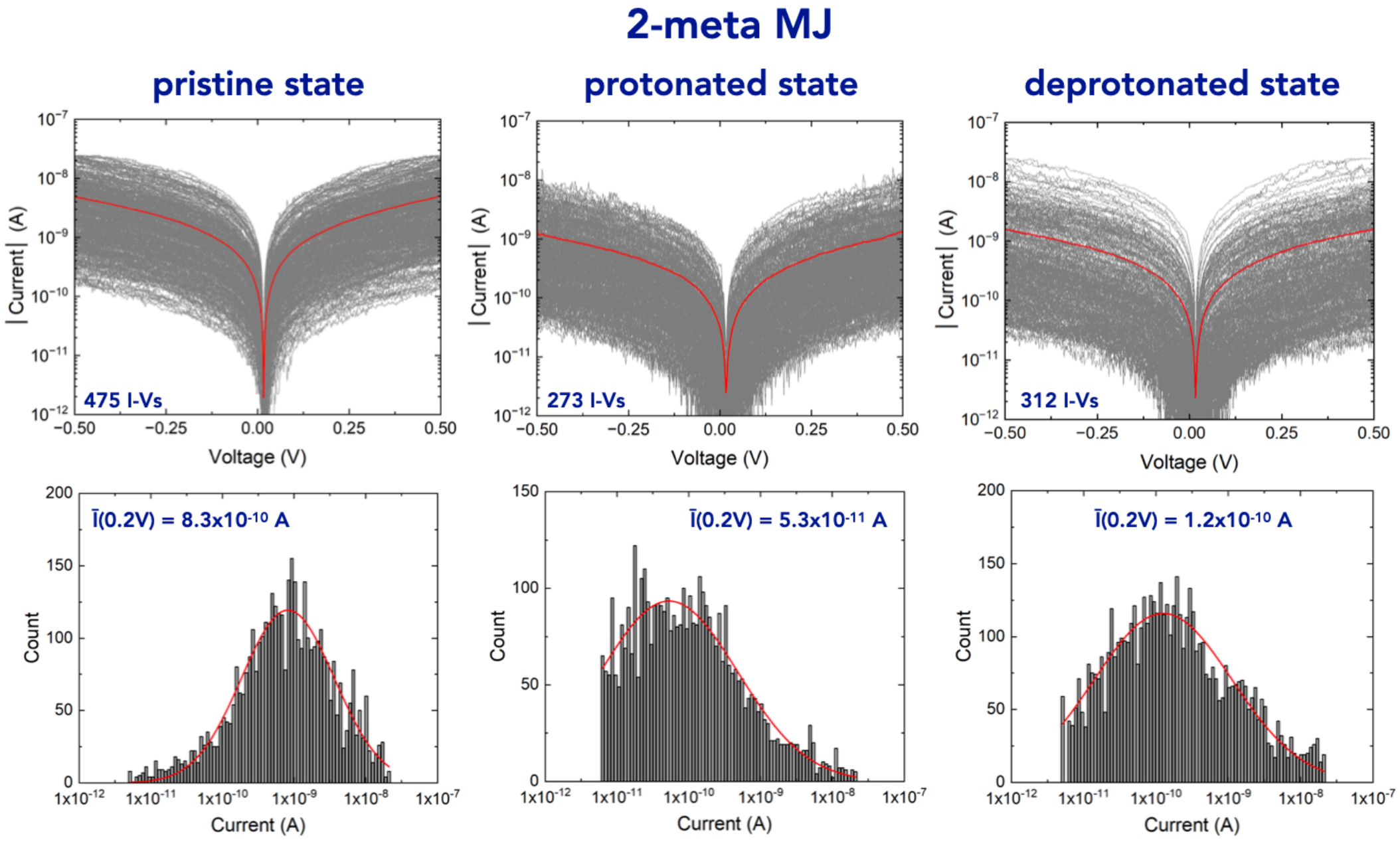

***Figure 3**. I-V datasets and corresponding histograms of the currents (at 0.2 V) for the **2-meta** MJs: pristine state, after protonation and deprotonation. In the datasets, the red lines are the mean $\bar{I}$-V curves. In the current histograms, the red lines are the fits by a log-normal distribution. The mean values of the current, $\bar{I}$, are indicated in the panels.*

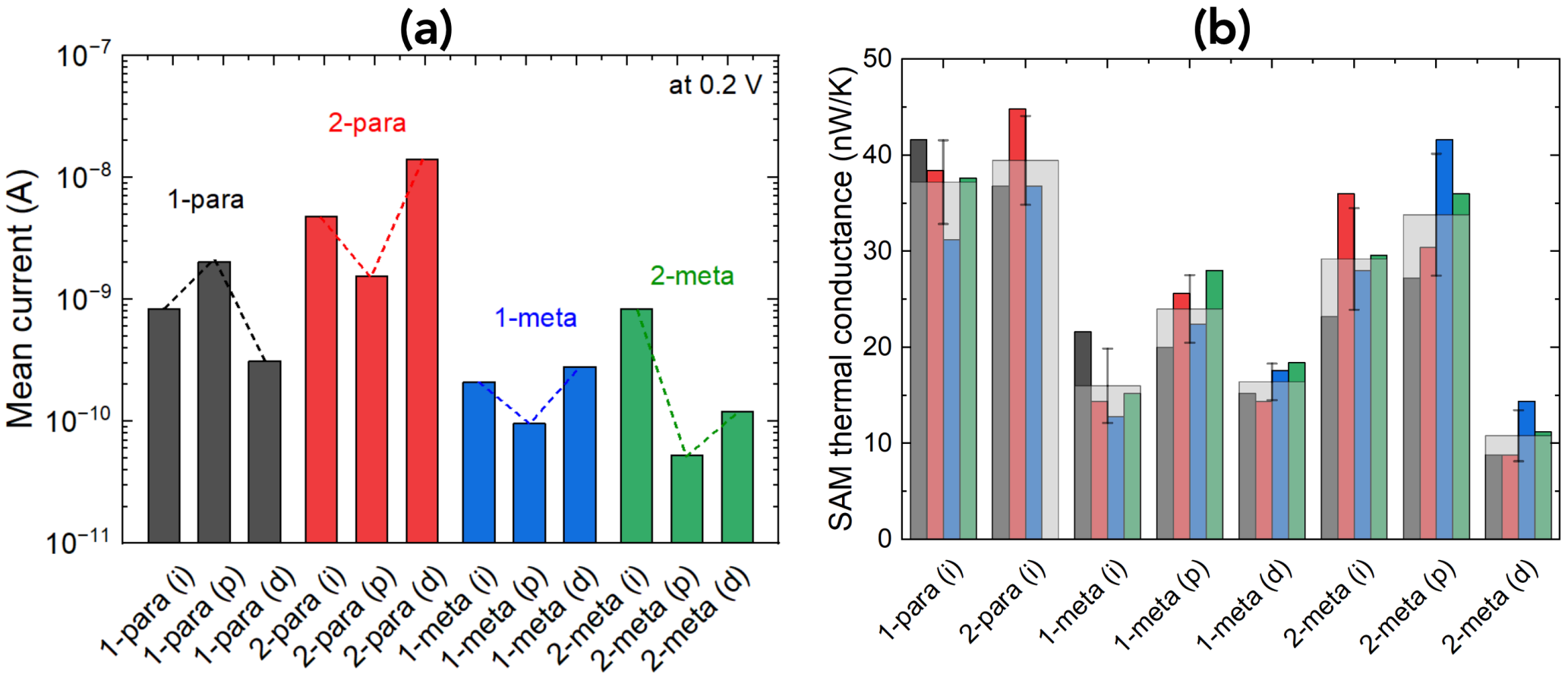

***Figure 4. (a)** Evolution of the mean currents (datasets in Figs. 2 and 3) at 0.2 V (data for **1-para** and **2-para** from Ref. 34): (i)=pristine, (p)=after protonation, (d) after deprotonation. **(b)** Evolution of the SAM thermal conductances. The color bars correspond to measurements on 3 - 4 different zones on the SAM. Light gray areas are the average values (with standard deviations as error bars). The values are summarized in Table S2 (supplementary information).*

***3.3. Thermal conductance (SThM)**.*

Figures 5-a and 5-b show typical NP-SThM measurements on the pristine **1-*meta*** and **2-*meta*** SAMs. The tip temperature is measured versus the vertical distance between the tip and the sample surface (z-displacement of the piezo actuator). From z=0 (the tip is far from the surface) $T_{tip}$ slightly decreases when approaching the tip due to the increase of radiative and convective heat conductions until a sudden temperature jump is observed when the tip is touching the surface. The amplitude of this temperature step ($T_{NC}$-$T_C$) is due to the addition of another thermal conduction pathway through the sample. Higher the amplitude of $T_{NC}$-$T_C$, higher the sample thermal conductance. From data in Figs. 5-a and 5-b, we can deduce that the thermal conductance of the **2-*meta*** SAM is larger than that of the **1-*meta*** SAM.

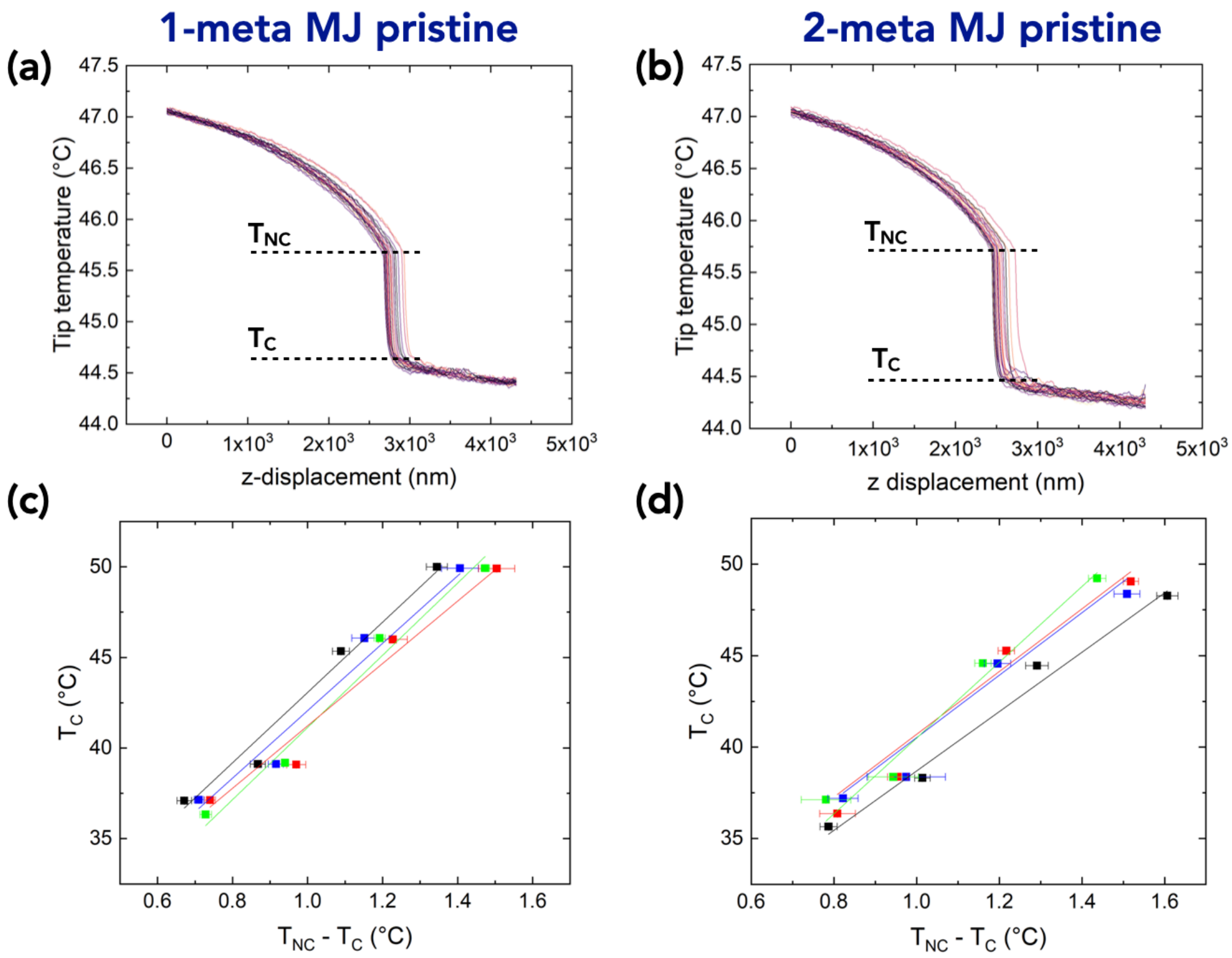

***Figure 5**. Typical NP-SThM measurements: $T_{tip}$ vs. z-displacement (25 traces) for **(a) 1-meta** pristine and **(b) 2-meta** pristine SAM. The dashed lines indicate the values of $T_{NC}$ and $T_C$. **(c-d)** $T_C$ vs. $T_{NC}$-$T_C$ plots at 4 different locations on the SAMs for the **1-meta** and **2-meta** SAMs, respectively. The lines are linear fits to calculate the sample thermal conductance $G_{th,SAM/Au}$.*

The thermal conductances are calculated by plotting $T_C$ vs. $T_{NC}$-$T_C$ (see details in the supplementary information and Ref. 41) from several $T_{tip}$-z curves recorded at increasing heat flux (by increasing the voltage supply of the Wheatstone bridge in which the resistive SThM tip is connected). The slope of this curve is inversely proportional to the thermal conductance (Eq. S1). Figs. 5-c and 5-d show these plots for the **1-*meta*** and **2-*meta*** MJs (in the pristine state, data for the protonated and deprotonated samples are shown in Fig. S19). These measurements were carried out at 3-4 places on the SAM surfaces, for which we calculated the mean thermal conductance. Since the SAMs are very thin, the measured thermal conductance is influenced by the high thermal conductance of the Au substrate and the measured thermal conductance is that of the Au/SAM system, $G_{th}$(SAM/Au). The thermal conductance of the SAM, $G_{th}$(SAM), is calculated by using the Dryden model that derived an analytical equation for the thermal conductance of two-layer samples (see Supplementary information, Eq. S2).[45] Figure 4-b summarizes the $G_{th}$(SAM) values for the **1-*meta*** and **2-*meta*** MJs in the pristine state, after protonation and deprotonation, and for the **1-*para*** and **2-*para*** MJs in the pristine state for comparison (data in Figs. 5, S19 and S20). The mean values of $G_{th}$(SAM) are summarized in Table S2. The thermal conductance of the SAMs (pristine) is almost the same for **1-*para*** and **2-*para*** (≈ 37-40 nW/K) and it is weaker for **1-*meta* and 2-*meta*** (≈ 16-29 nW/K). Protonation has a clear effect on the thermal conductivity of the **1-*meta***, with an increase to ≈ 24 nW/K, and a weaker effect on **2-*meta*** MJs (increase from 29 to 33.8 nW/K), in both cases reversible after deprotonation. We note that the increase in $G_{th}$(SAM) ($\delta G_{th}$ = 4.8 nW/K) for the ***2-meta*** MJ is slightly lower than the error bars (6.3-6.4 nW/K, see Table S2) and that this trend must be further confirmed by additional measurements to enlarge the statistical significance of this difference. A small value of 10.8 nW/K was observed for the deprotonaned **2-*meta*** MJs (*vide infra* for explanation).

## 4. DISCUSSION.

### *4.1. Energy diagrams of the MJs and electrical conductance variations*.

We fitted the I-V datasets with the analytical single energy level (SEL) model to determine the electronic energy diagram of the MJs. We extracted the energy position of the highest occupied molecular orbital (HOMO), $\varepsilon_H$ (with respect to the Fermi energy of the electrodes), $\Gamma_1$ and $\Gamma_2$ the molecule/electrode electronic coupling energies with the two electrodes (see details in the supplementary information, Eq. S7). Figure 6 gives the statistical distributions of $\varepsilon_H$ for the **1-*meta*** and **2-*meta*** MJs and the parameters $\Gamma_1$ and $\Gamma_2$ are shown in Fig. 7. Table 1 summarizes the mean values. The energetics are not strongly different between the **1-*meta*** and **2-*meta***, nor strongly affected by protonation/deprotonation. These results contrast with recent reports on benzimidazole-substituted terphenyl and bipyridine-based SAM MJs, showing a slight reduction of $\varepsilon_H$ by protonation.[46, 47] This feature also differs from that of the **1-*para*** and **2-*para*** MJs for which we previously reported that the electrical conductance increased for **1-*para*** upon protonation, while it decreased for **2-*para*** MJs, due to significant and different variations of the molecular orbitals and the position of the electrode Fermi energy for the two *para*-connected MJs (as inferred only from DFT calculations and not from a SEL model fitting).[34] Clearly, this behavior is not reproduced with the same molecules connected in the *meta* position. A possible reason would be Fermi level pinning in the *meta*-connected MJs due to the interaction between the π-orbitals and the metal electron wave functions,[48-50] which hinders the shift of the HOMO energy upon protonation. When the orbitals of the molecules overlap with the electron wave functions of metal electrons, a localized density of states (DoS) appears in the HOMO-LUMO band gap (referred to as chemisorption-induced gap states or metal-induced gap states).[50, 51] The Fermi level is pinned at the energy of these gap states so that to insure the charge neutrality between the filled and empty states in the gap, with electron transfers between the metal and the molecule and the creation of a metal-molecule interface dipole. If the gap state DoS is large enough (e.g., when the molecules are largely tilted or lying flat on the metal surface), any modification of the charge states at the metal/molecule interface results in a tiny shift of the Fermi level, or in other words, of the energy offset between the Fermi level and the molecular orbitals. Such a mechanism might account for the negligible variations of the measured HOMO position upon protonation/deprotonation in the *meta*-connected MJs. Such gap states in MJs have been theoretically and experimentally reported.[50-57] On the contrary, our previous report[34] showed that the opposite behavior in the electrical conductance variations of the ***1-para*** and ***2-para*** MJs is explained, in the absence of gap states as inferred by DFT calculations, by a large variation of the energy position of the Fermi level upon protonation. In any case, this calls for confirmation by UPS experiments. In this present case, the weak electrical conductance decrease upon protonation can be related to the observed variation of the SAM thickness (*vide supra,* ellipsometry measurements). This is due to the exponential decrease in the tunneling electron transmission function with increasing SAM thickness, which, in the SEL model, is reflected by the decrease in the $\Gamma_1$ and $\Gamma_2$ values (Fig. 7 and Table 1). It is known that this parameter decreases with the SAM thickness in the MJs.[58-60] The SAMs with the *meta*-connected molecules are less densely packed. In the pristine state, the molecules are initially largely tilted (≈ 66° from the surface normal, see details in the supplementary information) and it is likely that they move toward an upright position to accommodate the protons (intercalation) and then relax to the initial structure when the protons are removed. Thus the behavior of the SAMs with the *meta*-connected molecules is dominated by structural changes rather than by the intrinsic properties of the molecules. Note that the primary role of Γ to explain the changes in conductance of protonated MJs was also reported for other MJs (an increase in Γ and electrical conductance of protonated benzimidazone-substituted terphenyl MJs,[46] and a decrease in Γ and electrical conductance for protonated bipyridine MJs[47]). The lower electrical conductances for the **1-*meta*** and **2-*meta*** MJs (with respect to their counterparts in the *para* conformation) are also consistent with destructive quantum interferences,[1-3, 5, 6, 8, 11, 44] However, we cannot exclude a decrease in conductance induced by a lower molecule/electrode electronic coupling, since the large tilt can favor a physisorbed contact between the C-AFM tip and the side of the molecule rather than a chemisorbed S-Au contact when the molecules are standing more upright on the surface (as in the *para* conformation). Another reason would also be that, in the less packed SAMs, fewer molecules are contacted under the C-AFM tip.

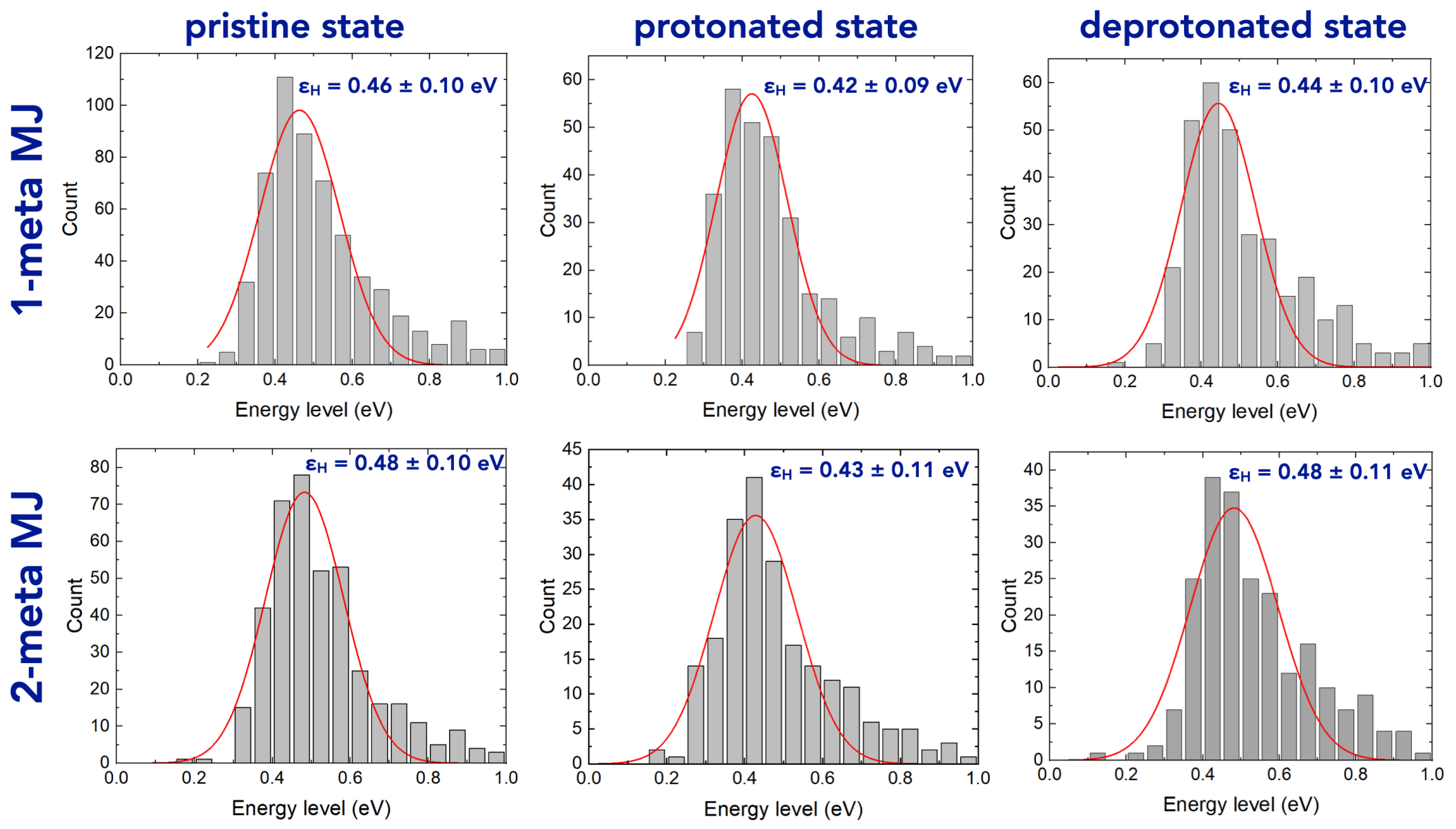

***Figure 6***. *Statistic distributions of the HOMO energy level ($\varepsilon_H$) with respect to the electrode Fermi energy as deduced from the analysis of the I-V datasets (Figs. 2 and 3) by the single energy level (SEL) model. The red lines are the fits by Gaussian distributions, the mean values (± standard deviations) are indicated in the panels and summarized in Table 1.*

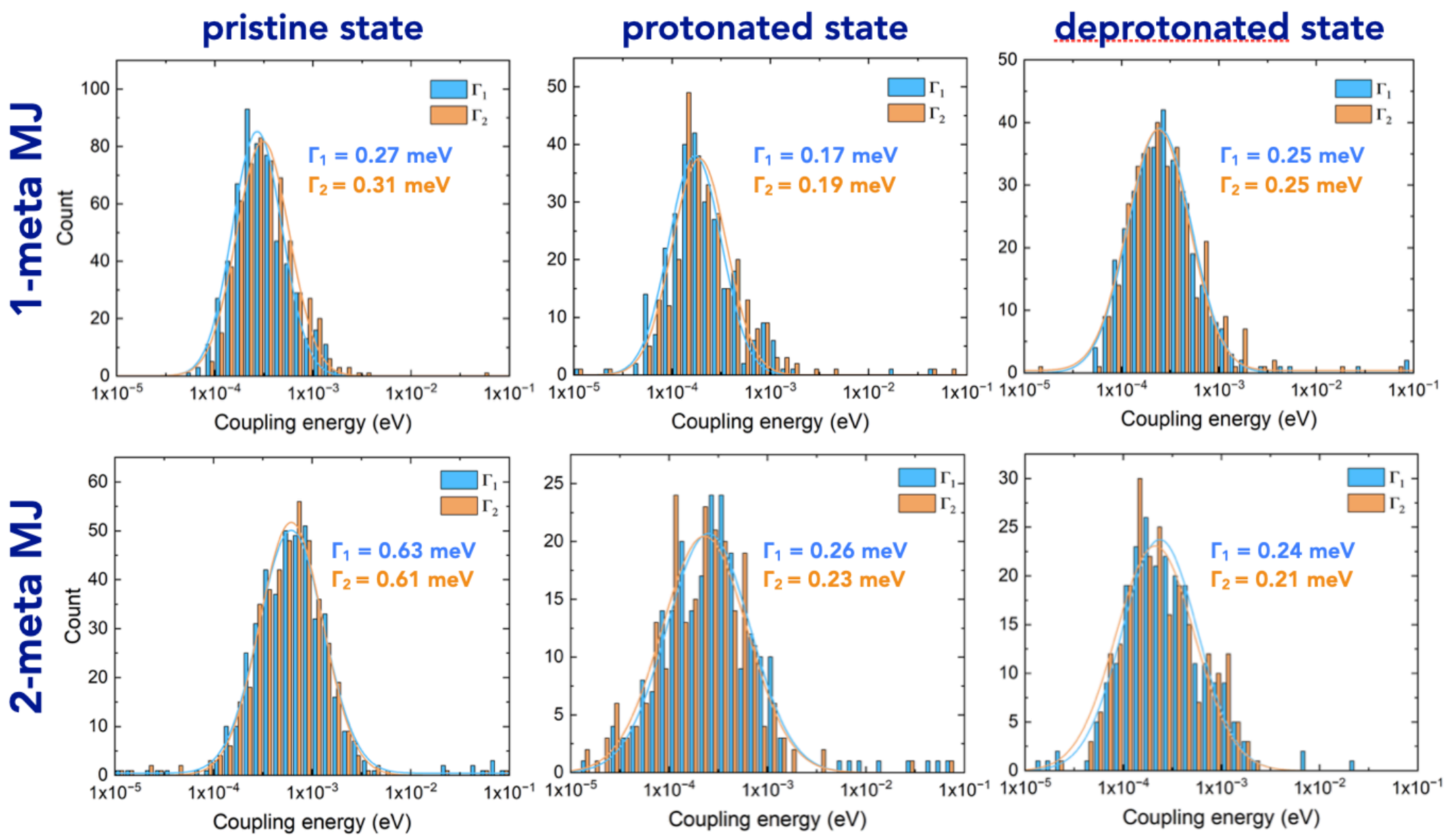

***Figure 7**. Statistical distributions of the molecule/electrode coupling energy for the **1-meta** and **2-meta** MJs in the 3 states. Solid lines are fits with a log-normal distribution and the means $\Gamma_1$ and $\Gamma_2$ are reported in Table 1.*

| | | pristine | protonated | deprotonated |
|---|---|---|---|---|
| **1-*meta* MJ** | **$\varepsilon_H$ (eV)** | 0.46 ± 0.10 | 0.42 ± 0.09 | 0.44 ± 0.10 |
| | **$\Gamma_1$ ; $\Gamma_2$ (meV)** | 0.27; 0.31 | 0.17; 0.19 | 0.25; 0.25 |
| **2-*meta* MJ** | **$\varepsilon_H$ (eV)** | 0.48 ± 0.10 | 0.43 ± 0.11 | 0.48 ± 0.11 |
| | **$\Gamma_1$ ; $\Gamma_2$ (meV)** | 0.63; 0.61 | 0.26; 0.23 | 0.24; 0.21 |

***Table 1**. Mean values (± standard deviations) of the HOMO energy level $\varepsilon_H$ (data Fig. 6) and electrode coupling energy $\Gamma_1$ and $\Gamma_2$ (data Fig. 7) for the **1-meta** and **2-meta** MJs in the three states.*

***4.2. Thermal conductance**.*

Since the thermal conductances of these SAMs of benzo-bis(imidazole) molecules (whether in *para* or *meta* conformations) were not reported before (to the best of our knowledge), the first comment concerns the measured values (in the pristine state), $G_{th}$(SAM) ≈ 16 - 40 nW/K (or ≈ 16-40 $MWm^{-2}K^{-1}$, for the purpose of comparison between the various techniques: SThM, 3ω, thermoreflectance, which used samples of quite different sizes). These values are in the same range (typically ≈ 10-70 $MWm^{-2}K^{-1}$)[61] as the previously reported values for SAMs of other molecules, e.g., alkylthiols and alkyldithiols of various chain lengths (up to 18 carbon atoms),[62-65] BTBT derivatives,[66] and organometallic molecules.[27] We first discuss the modification of the thermal conductances of the *meta*-connected SAMs by the protonation/deprotonation, a feature not experimentally, nor theoretically, reported before. The thermal conductance is primarily controlled by the vibration modes of the molecules. It is unlikely that the addition of light protons can significantly modify the vibrational spectra of the molecules. A significant reduction in the thermal conductance of MJs has been theoretically predicted for substitutions by heavier atoms.[31, 67, 68] Thus, we assume that the intrinsic thermal conductance should remain unaffected by the protonation. We make the hypothesis that the variations of the thermal conductances of **1-*meta*** and **2-*meta*** MJs under protonation and deprotonation is likely related to the modification of the structural organization of the molecules in the SAMs as reflected by the changes in the SAM thickness. As spectroscopic ellipsometry is a well-established method to measure SAM thickness down to 0.5 nm,[69-71] this hypothesis is plausible, albeit further confirmation using infrared spectroscopy (grazing-angle FTIR or PM-IRRAS) would be required to provide more direct information on the molecular orientation and structural organization of the SAM. In particular, the measured $G_{th}$(SAM) is made of the contributions of the molecule core and two interfacial Kapitza resistances at the bottom Au and top SThM tip electrodes.[72-75] It is known that the physical and chemical nature of the electrode-molecule interfaces strongly modulate the thermal conductance for the same molecule. For instance, the thermal conductance of alkyl-based MJs is larger for a chemisorbed molecule-top electrode interface (-SH terminated chain) than for a van der Waals interface (*e.g.*, -$CH_3$ terminated chain) by a factor ≈ 2.[63] The thermal conductance of MJs with chemisorbed interfaces at the two electrodes (alkanedithiols) is higher than that with only one chemisorbed interface (alkanemonothiols).[65, 76] The chemical nature of other end groups (Br, $NH_2$, SH) also has a slight effect ($G_{th,Au} > G_{th,Br} > G_{th,NH2} > G_{th,CH3}$),[63] as well as the nature of the metal electrode (Au, Ag, Pt, Pd).[65] When the **1-*meta*** molecules in the SAM straighten up, the conformation of the bottom Au-S-C- link is modified, which can induce a modification of the interfacial thermal conductance. At the top side of the SAMs, more SH groups become exposed to the SAM surface that can allow the formation of a chemisorbed -C-S-Pd interface with the SThM tip, rather than a "mechanical" contact between the tip and the side of the benzo-bis(imidazole) molecules, leading to an increase in the thermal conductance,[4] as observed in our experiments (Fig. 4-a). After deprotonation, the molecules returned to the large tilted configuration and the thermal conductance recovers its initial value. The same trends are observed for the **2-*meta*** MJs, albeit with a weaker amplitude (it is likely that the bulkier side groups may complicate this simple hypothetical scheme). The low thermal conductance after deprotonation in that case (≈ 11 nW/K) might come from a partial degradation of the SAMs, since a second protonation/deprotonation cycle leads to weaker changes in the thermal conductance, while it is almost reproduced for ***1-meta*** MJs (Fig. S21).

Finally, we discuss the observed feature that $G_{th}$ (in the pristine state) for the *meta*-connected SAMs are lower than those in the *para*-connected SAMs (≈ 16-29 nW/K and ≈ 37-40 nW/K, respectively, Fig. 4-b), in agreement with the recent experiment by Yelishala and coworkers,[33] and in agreement with theoretical estimates that the phonon interference effects are responsible for the lower thermal conductances of MJs with *meta*-connected molecules.[28-30] However, these latter experimental results were obtained on single molecule MJs. In our case, we cannot exclude that intermolecular coupling and substrate-mediated interactions in the SAMs may also modify these interference effects and that the structural differences (molecular packing density, tilt angle) between the *meta*- and *para*-connected SAMs can also contribute to this difference in the thermal conductance. In addition, we note that the variations of the interface thermal conductance can also play a role. For instance, it was also calculated that the thermal conductance of *meta*-connected MJs can be larger than of *para*-connected ones when the sulfur anchor atom of the molecule is connected to an atop Au atom (instead of a hollow Au binding, for which the thermal conductance of *para*-connected MJ is larger than the *meta*-connected MJ).[32] Here, since both the ***1-para*** and ***2-para*** molecules are nearly perpendicular (SAM thickness ≈ length of the molecule, as inferred from ellipsometry, see supplementary information), it is likely that S-Au bonds can be more easily formed at the bottom and top electrodes (compared to largely tilted molecules in the *meta*-connected MJs), thus partly contributing to the observed larger thermal conductances (Fig. 4-b) than for the MJs with *meta* connectivity (for example, alkyldithiol MJs have a slightly higher thermal conductance than alkylthiol MJs).[64, 77] A last issue concerns the role of the $Cl^-$ counterions present in the SAMs to maintain charge neutrality. Since the $Cl^-$ is heavier than the proton, it can likely alter the vibration modes of the molecules in the SAMs, and contribute to the modification of the thermal conductance. For instance, the theoretical thermal conductance of Cl-substituted biphenyl molecule is reduced by a factor ≈ 2 compared to the unsubstituted molecule.[67] Here, the $Cl^-$ ions are just inserted into the SAMs, but a similar effect (albeit with a weaker amplitude than in the case of substitution) can be supposed. Since the thermal conductance of the protonated *meta*-connected SAMs increases, the role of the $Cl^-$ counterions would be to partly mask the neat increase due to the protonation alone. Similarly, the protonation of the molecules and the presence of $Cl^-$ counterions can increase the refractive index due to the presence of the charged species ($NH_3^+$, $Cl^-$) and the resulting presence of dipoles in the SAM (higher polarizability). This refractive index increases would lead to a decrease in the deduced thickness. Neglecting this effect in the ellipsometry analysis using a constant refractive index (see the Supporting Information) would also contribute to underestimating the thickness of the protonated SAMs. Thus, this effect cannot explain the observed thickness increase of the protonated *meta*-connected SAMs. To verify whether the presence of $Cl^-$ affects the thermal conductance and to what extent, further experiments could be required using chlorine-free acid, as reported by several groups, e.g., trifluoroacetic,[78-80] trifluoromethanesulfonic[78] or tetrafluoroboric[46, 47] acids in solution.

Finally, we discuss further challenges and practical thermoelectric device perspectives.

The thermoelectric factor of merit ZT is given by $G_{el}S^2T/G_{th}$, with $G_{el}$ the electrical conductance, $G_{th}$ the thermal conductance, S the Seebeck coefficient and T the temperature. Irrespective of S, to improve ZT, we have to increase the $G_{el}/G_{th}$ ratio. We estimated these ratios for the series of MJs measured in this study (Fig. S22 in the Supporting Information). We conclude that the molecule ***2-para*** MJ seems the most promising with the highest $G_{el}/G_{th}$ ratio. The protonation degrades these ratios (Fig. S22). To completely evaluate the thermoelectric performances of these MJs, the Seebeck coefficient must be accurately measured. However, considering an average value S ≈ 10 µV/K (from a compilation of many experimental values, as reported in several review papers),[61, 81-83] a very low ZT value can be predicted (≈ 3-4x$10^{-6}$) at room temperature, far away from the required ZT>1 for practical applications. Albeit the SAM thermal conductances $G_{th}$ (here between 20 and 40 nW/K, Fig. 4-b) are on a par with the values reported for SAMs based on other molecules (see section 4.2), further improvements are mandatory, for example (but not limited to), exploring other chemical groups at the molecule/ electrode interfaces, which are known to tune the electrical and thermal conductances.[56, 63]

A practical thermoelectric device requires two macroscopic electrodes (i.e., not a C-AFM or SThM tip, nor a macroscopic Hg or eGaIn drop). However, it is challenging to form a top macroscopic electrode on SAMs. For example, vacuum evaporation of a metal electrode on top of a SAM very frequently results in a poor fabrication yield due to metallic filaments through the SAMs (for a review on this topic and possible solutions, see Ref. 84, 85). Moreover, in such a large-area vertical MJs, access of protons to the SAM is largely hindered. A more powerful approach is based on lateral 2D networks of metal nanoparticles functionalized with the molecules of interest and connected to their periphery by standard lithographed electrodes. This device has been proven a versatile platform to study the basic electronic properties of molecular electronics,[86-90] as well as to evaluate potential applications, such as chemical or strain sensors, plasmonic devices,[89] reconfigurable logic gates,[91-95] and neuro-inspired reservoir computing.[95-98] It is likely that this platform would also be prone to thermoelectric devices.

Due to the low packing density of the meta-connected MJs, it is difficult to conclude whether the changes in the thermal conductance is an intrinsic property of the molecules or whether it is induced by the modifications of the structural organization of the molecules in SAMs, which impact the thermal conductance at the SAM/C-AFM tip interface (or a combination of the two). This is due to the fact that when the low density packed meta-connected SAMs are exposed to the acid or base vapors, the molecules are anchored only at one end on the bottom Au surface and the molecules in the molecules in the SAMs can undergo some structural reorganizations. To disentangle the intrinsic and extrinsic effects, a first solution would be to increase the molecular packing. However, to give ourselves the best possible chance of achieving the highest possible density, we have chosen a very long time of immersion for the SAM formation and a largely in excess concentration of molecules in the solution (3 days and 1 mM, see supporting information). A para-substitution of the thiol anchoring group generally leads to a more linear molecular/anchoring geometry, which can favor an efficient lateral packing in the SAM. A meta-position of the thiol group at ca. 120° with respect to the long axis of the molecule backbone can generally induce a larger tilt of the molecule, a larger effective molecular surface area and a greater hindrance to molecular packing.[99] A likely better approach to limit these structural variations would be to anchor the molecules at their two ends between adjacent Au nanoparticles in the 2D molecule/nanoparticle platform discussed above, which provide better structural stability.[100] A better stability would also likely improve the number of protonation/deprotonation cycles of the devices.

## 5. CONCLUSION.

This work highlights the behaviors of the thermal and electronic conductances of SAM-based molecular junctions made of two benzo-bis(imidazole) derivatives upon the protonation/deprotonation of imidazole moieties, when the molecules are connected to the electrodes by thiol anchoring groups in the *meta*-conformation. We confirmed that the thermal conductance in the *meta*-conformation is lower than in the *para*-conformation, as theoretically suggested due to phonon destructive quantum interferences[28, 29] and as recently experimentally reported for oligo(phenylene ethynylene)-based molecular junctions.[33] The reversible changes in the thermal and electronic conductances of the *meta*-conformation molecular junctions upon protonation/

deprotonation of the imidazole core in the molecules are likely related to changes in the molecule organization in the self-assembled monolayers.

## AUTHOR CONTRIBUTIONS.

S.G-C. prepared the SAMs and measured the samples (ellipsometry, C-AFM, SThM). S.P. and O.S. synthesized the molecules. S.G-C. and D.V. analyzed the results. D.V. wrote the manuscript with contributions and comments from all the authors. All the authors have approved the manuscript.

## CONFLICTS OF INTEREST.

The authors declare no competing financial interest.

## DATA AVAILABILITY.

The data supporting this findings are given in the main article and as part of the Supplementary Information. Supporting information: Molecule synthesis and characterization, SAM preparation and characterization, C-AFM and SThM experimental protocols, C-AFM data analysis with the single energy level model, additional figures.

## ACKNOWLEDGEMENTS.

We thank the financial support from ANR (# ANR-21-CE30-0065). We acknowledge K. Kondratenko for preliminary SThM measurements on the *para*-connected MJs and D. Guérin for his help and advice on the fabrication of the SAMs. S.P. and O.S. thank the CNRS, Nantes University and Aix-Marseille University for their support.

## REFERENCES.

**ToC graphic.**

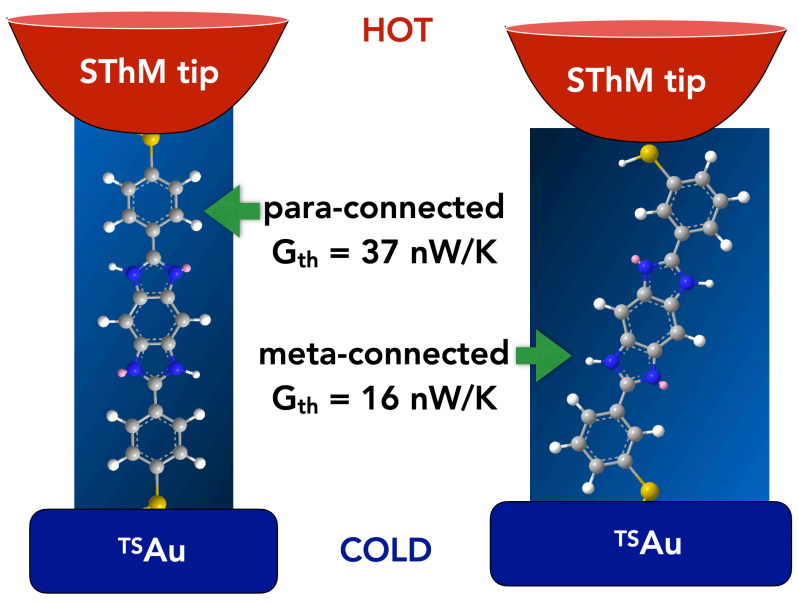

We report a lower thermal conductance of benzo-bis(imidazole) molecular junctions connected in meta conformation. The effects of the protonation/deprotonation on the electrical and thermal conductances are also investigated.

# Benzo-bis(imidazole) Self-Assembled Monolayers Molecular Junctions in *meta* or *para* Conformation: Effects of Protonation on the Electrical and Thermal Conductances.

Sergio Gonzalez-Casal,[1,§] Simon Pascal,[2,3] Olivier Siri[2] and Dominique Vuillaume.[1*]

*1) Institute for Electronics Microelectronics and Nanotechnology (IEMN), CNRS, Villeneuve d'Ascq, France.*
*2) Centre Interdisciplinaire de Nanoscience de Marseille (CINaM), CNRS, Luminy, France.*
*3) Chimie et Interdisciplinarité, Synthèse, Analyse, Modélisation( CEISAM), CNRS, Nantes, France.*

*§ Now at Institut Néel, Université Grenoble Alpes, Grenoble, France.*

** Corresponding author: dominique.vuillaume@iemn.fr*

## Supplementary Information.

## I. Synthesis.

***Analytical methods and apparatus.***

Melting points (M.P.) were measured in open capillary tubes with a STUART SMP30 melting point apparatus and are uncorrected. Nuclear magnetic resonance (NMR) spectra were recorded on a JEOL ECS400 NMR spectrometer at room temperature (otherwise noted). NMR chemical shifts are given in ppm (δ) relative to $Me_4Si$ with solvent resonances used as internal standards ($CDCl_3$: 7.26 ppm for $^1H$ and 77.2 ppm for $^{13}C$; DMSO-$d_6$: 2.50 ppm for $^1H$ and 39.5 ppm for $^{13}C$). Fourier-transform infrared spectroscopy (FTIR) spectra were recorded on an Agilent Cary 630 FTIR equipped with an attenuated total reflectance (ATR) sampling. HRMS analyses were performed on a QStar Elite (Applied Biosystems SCIEX) or a SYNAPT G2 HDMS (Waters) spectrometers by the “Spectropole” of the Aix-Marseille University. These two instruments are equipped with an ESI or MALDI source, and a TOF mass analyzer.

***Compound 1 : (2-((3-bromophenyl)thio)ethyl)trimethylsilane.***

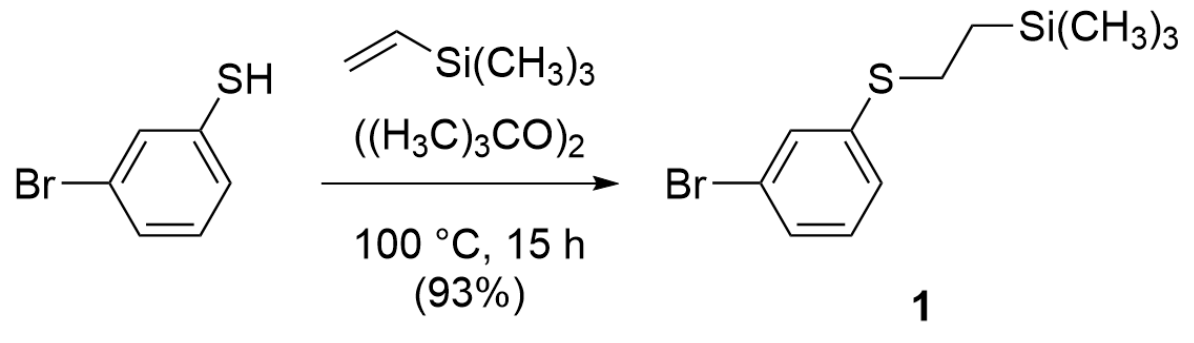

3-Bromothiophenol (3 g, 1.64 mL, 15.87 mmol, 1 equiv.), vinyltrimethylsilane (2.33 mL, 15.87 mmol, 1 equiv.) and di-*tert*-butyl peroxide (.41 mL, 2.22 mmol, 0.14 equiv.) were mixed in a dry Schlenk tube under argon atmosphere. The mixture was heated to 100 °C for 15 hours then cooled to room temperature, diluted with petroleum ether (*ca*. 100 mL) and washed with an aqueous NaOH solution (1 M). The aqueous layer was extracted three times with petroleum ether and the combined organic layers were dried over anhydrous sodium sulfate, filtered on a paper and concentrated under reduced pressure. Purification of the oily residue by silicagel flash column chromatography using petroleum ether as eluent afforded the pure product **1** as a colorless oil in 93% yield (4.285 g, 14.81 mmol).

**$R_f$:** 0.41 ($SiO_2$, petroleum ether). **$^1$H NMR ($CDCl_3$, 400 MHz):** δ = 7.40 (d, $^4J$ = 1.2 Hz, 1H, CH), 7.27 (d, $^3J$ = 7.8 Hz, 1H, CH), 7.19 (d, $^3J$ = 7.8 Hz, 1H, CH), 7.13 (d, $^3J$ = 1.2 Hz, 1H, CH), 2.95 (m, 2H, $SCH_2$), 0.93 (m, 2H, $CH_2$), 0.05 (s, 9H, $Si(CH_3)_3$). **$^{13}$C{$^1$H} NMR ($CDCl_3$, 100 MHz):** 140.1 (C), 130.9 (CH), 130.2 (CH), 128.6 (CH), 127.0 (CH), 122.9 (C), 29.4 ($SCH_2$), 16.8 ($CH_2$), 1.6 ($Si(CH_3)_3$). **FTIR (neat, $cm^1$):** ν = 3055, 2952, 2896, 1574, 1557, 1458, 1400, 1248, 1068, 1010, 833, 751, 677. **HRMS (ESI+)** calculated for the second isotopic peak of $[M+Na]^+$: 312.9875 ($C_{11}H_{17}BrSSiNa^+$), found: 312.9877 (the maximum value corresponds to the second isotopic peak).

**Compound 2 : 3-((2-(trimethylsilyl)ethyl)thio)benzaldehyde.**

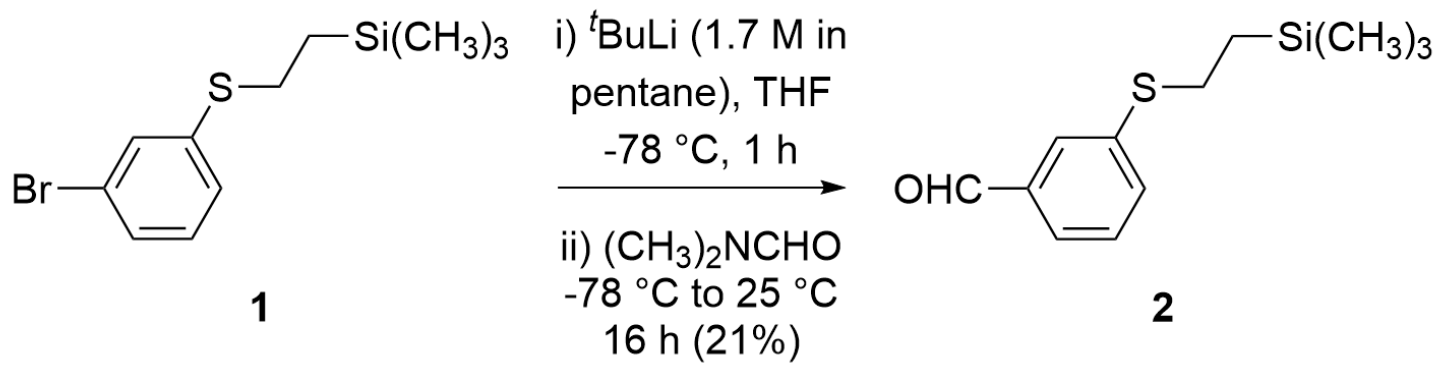

Compound **1** (1 g, 3.46 mmol, 1 equiv.) was dissolved in 25 mL of freshly distilled tetrahydrofuran and the solution was cooled to 78 °C using liquid nitrogen in an acetone bath and a 1.7 M solution of *tert*-butyllithium in pentane (2.44 mL, 4.15 mmol, 1.2 equiv.) was added dropwise. The solution was stirred for 1 h at 78 °C and was added dropwise by anhydrous *N*,*N*-dimethylformamide (0.32 mL, 4.15 mmol, 1.2 equiv.), stirred at this temperature for 1 h, then allowed to warm to room temperature (25 °C) overnight (15 h). The reaction mixture was poured into an aqueous saturated solution of $NH_4Cl$ (*ca*. 100 mL) and was extracted several times with dichloromethane. The combined organic layers were dried over anhydrous sodium sulfate, filtered on a paper and concentrated under reduced pressure. Purification of the oily residue by silicagel flash column chromatography using petroleum ether/dichloromethane (7:3 then 1:1) as eluent afforded the pure product **2** as a colorless oil in 21% yield (174 mg, 0.73 mmol).

**$R_f$:** 0.39 ($SiO_2$, petroleum ether/dichloromethane, 7:3). **$^1$H NMR ($CDCl_3$, 400 MHz):** δ = 9.98 (s, 1H, CHO), 7.76 (s, 1H, CH), 7.64 (d, $^3J$ = 7.5 Hz, 1H, CH), 7.52 (d, $^3J$ = 7.5 Hz, 1H, CH), 7.44 (d, $^3J$ = 7.5 Hz, 1H, CH), 3.02 (m, 2H, $SCH_2$), 0.95 (m, 2H, $CH_2$), 0.06 (s, 9H, $Si(CH_3)_3$). **$^{13}$C{$^1$H} NMR ($CDCl_3$, 100 MHz):** 192.1 (CHO), 139.6 (C), 137.0 (C), 134.2 (CH), 129.5 (CH), 128.6 (CH), 127.2 (CH), 29.2 ($SCH_2$), 16.7 ($CH_2$), 1.6 ($Si(CH_3)_3$). **FTIR (neat, cm$^1$):** ν = 3385, 3058, 2953, 2898, 2821, 2723, 2120, 1698, 1573, 1469, 1422, 1381, 1249, 1197, 1164, 1096, 1010, 834, 783, 756, 726, 683. **HRMS (ESI+)** calculated for $[M+Na]^+$: 261.0740 ($C_{20}H_{17}N_6^+$), found: 261.0741.

***Compound 1-meta: 2-(3-((2-(trimethylsilyl)ethyl)thio)phenyl)-6-(3-(((trimethylsilyl)methyl)thio)phenyl)-1,5-dihydrobenzo[1,2-d:4,5-d']diimidazole.***

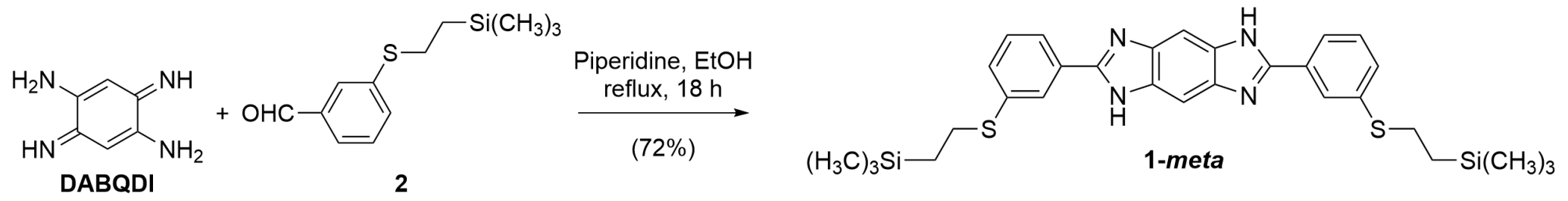

2,5-diamino-1,4-benzoquinonediimine (50 mg, 0.37 mmol, 1 equiv.) and compound **2** (184 mg, 0.77 mmol, 2.1 equiv.) were mixed in 4 mL of absolute ethanol. Three drops of piperidine were added and the solution was heated to reflux overnight (18 h). After it was cooled down to room temperature and stored for ten minutes in a fridge, the reaction mixture was filtered on a sintered glass funnel (porosity 4) and the isolated precipitate was washed consecutively with small quantities of ethanol, diethyl ether, pentane and finally dried under vacuum to afford the product **1-*meta*** as a beige powder in 72% yield (122 mg, 0.21 mmol).

**M.P.:** 324-326 °C. **$^1$H NMR (DMSO-*d$_6$*, 400 MHz, 353 K):** δ = 8.15 (s, 2H, CH), 8.01 (d, $^3J$ = 7.5 Hz, 4H, CH), 7.70 (s, 2H, CH), 7.49 (t, $^3J$ = 7.5 Hz, 4H, CH), 7.42 (d, $^3J$ = 7.5 Hz, 4H, CH), 3.12 (t, $^3J$ = 8.2 Hz, 4H, $SCH_2$), 0.98 (t, $^3J$ = 8.2 Hz, 4H, $CH_2$), 0.08 (s, 18H, $Si(CH_3)_3$). **$^{13}C\{^1H\}$ NMR (DMSO-*d$_6$*, 100 MHz):** 151.3 (C), 141.8 (C), 137.9 (C), 132.6 (C), 131.0 (C), 129.6 (CH), 128.9 (CH), 125.5 (CH), 123.4 (CH), 98.9 (CH), 28.2 ($SCH_2$), 16.2 ($CH_2$), 1.6 ($Si(CH_3)_3$). **FTIR (neat, cm$^1$):** ν = 3740, 3035, 2948, 2915, 2846, 2808, 2724, 2098, 1642, 1596, 1555, 1473, 1436, 1401, 1368, 1293, 1244, 1160, 1093, 1009, 961, 888, 833, 794, 754, 709, 685. **HRMS (ESI+)** calculated for $[M+H]^+$: 575.2149 ($C_{30}H_{39}N_4S_2Si_2^+$), found: 575.2150.

***Compound 2-meta: 4,4'-(2-(3-((2-(trimethylsilyl)ethyl)thio)phenyl)-6-(3-(((trimethylsilyl)methyl)thio)phenyl)benzo[1,2-d:4,5-d']diimidazole-1,5-diyl)dianiline.***

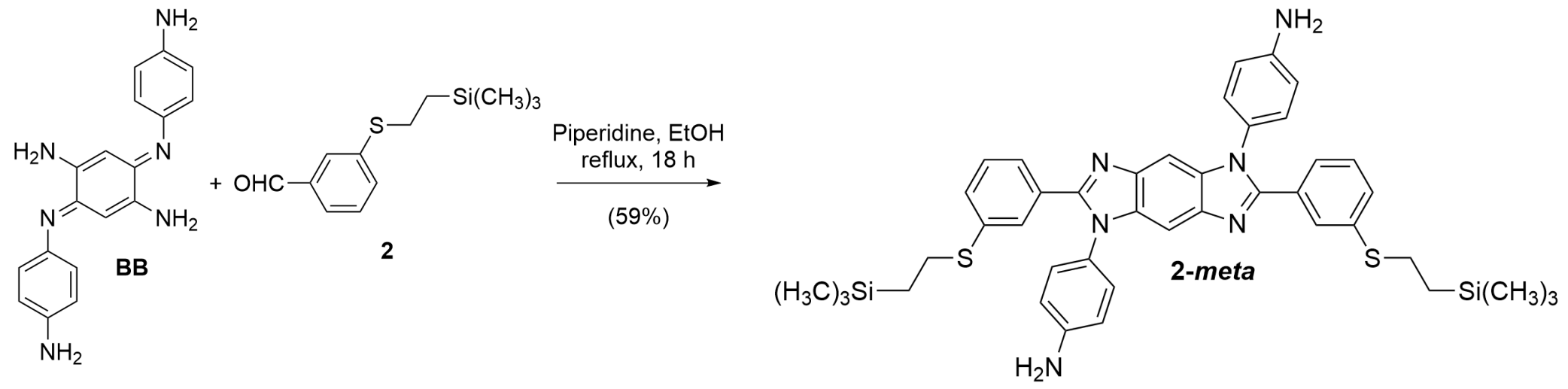

Bandrowski's base (100 mg, 0.31 mmol, 1 equiv.) and compound **2** (157 mg, 0.66 mmol, 2.1 equiv.) were mixed in 3 mL of absolute ethanol. Three drops of piperidine were added and the solution was heated to reflux overnight (18 h). After it was cooled down to room temperature and stored for ten minutes in a fridge, the reaction mixture was filtered on a sintered glass funnel (porosity 4) and the isolated precipitate was washed consecutively with small quantities of ethanol, diethyl ether and pentane; then finally dried under vacuum to afford the product **2-*meta*** as a beige powder in 59% yield (140 mg, 0.18 mmol).

**M.P.:** 272-274 °C. **$^1$H NMR (DMSO-*d$_6$*, 400 MHz):** δ = 7.51 (s, 2H, CH), 7.39 (m, 2H, CH), 7.30 (m, 6H, CH), 7.08 (d, $^3J$ = 7.5 Hz, 4H, CH), 6.73 (d, $^3J$ = 7.5 Hz, 4H, CH), 5.52 (s, 4H, $NH_2$), 2.89 (m, 4H, $SCH_2$), 0.83 (m, 4H, $CH_2$), 0.04 (s, 18H, $Si(CH_3)_3$). **$^{13}C\{^1H\}$ NMR (DMSO-*d$_6$*, 100 MHz):** 152.2 (C), 149.2 (C), 140.1 (C), 137.3 (C), 136.1 (C), 131.0 (CH), 128.8 (CH), 128.3 (CH), 128.1 (CH), 127.6 (CH), 125.8 (CH), 124.4 (C), 114.4 (CH), 27.8 ($SCH_2$), 15.9 ($CH_2$), 1.6 ($Si(CH_3)_3$). **FTIR (neat, cm$^1$):** ν = 3740, 3468, 3365, 3321, 3197, 2947, 2890, 1630, 1515, 1473, 1438, 1393, 1360, 1303, 1245, 1210, 1183, 1110, 1079, 1010, 978, 909, 839, 830, 797, 756, 709, 690. **HRMS (ESI+)** calculated for $[M+H]^+$: 757.2993 ($C_{42}H_{49}N_6S_2Si_2^+$), found: 757.2993.

***NMR spectra.***

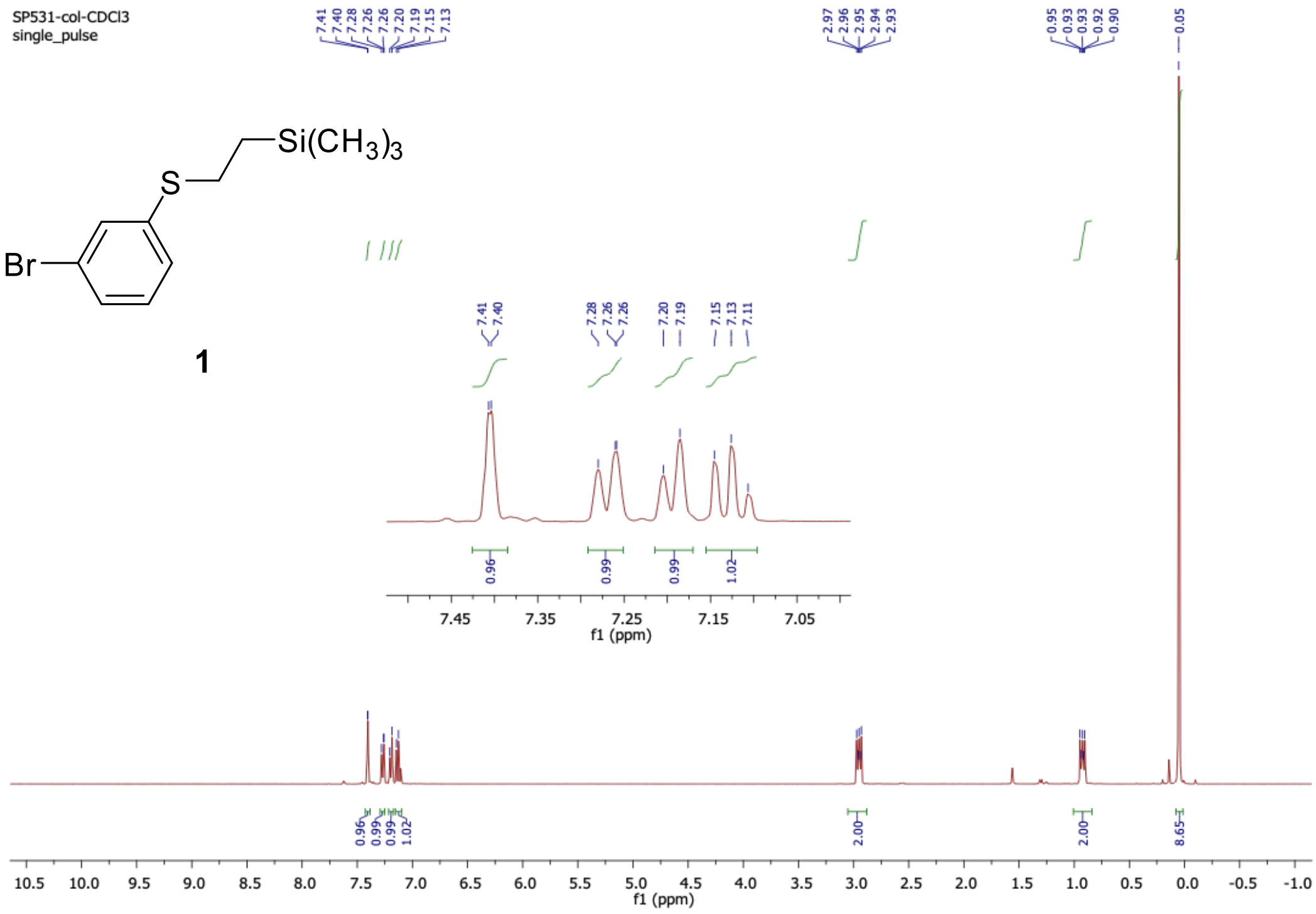

***Figure S1**. $^{1}H$ NMR (400 MHz, $CDCl_3$) of compound **1**.*

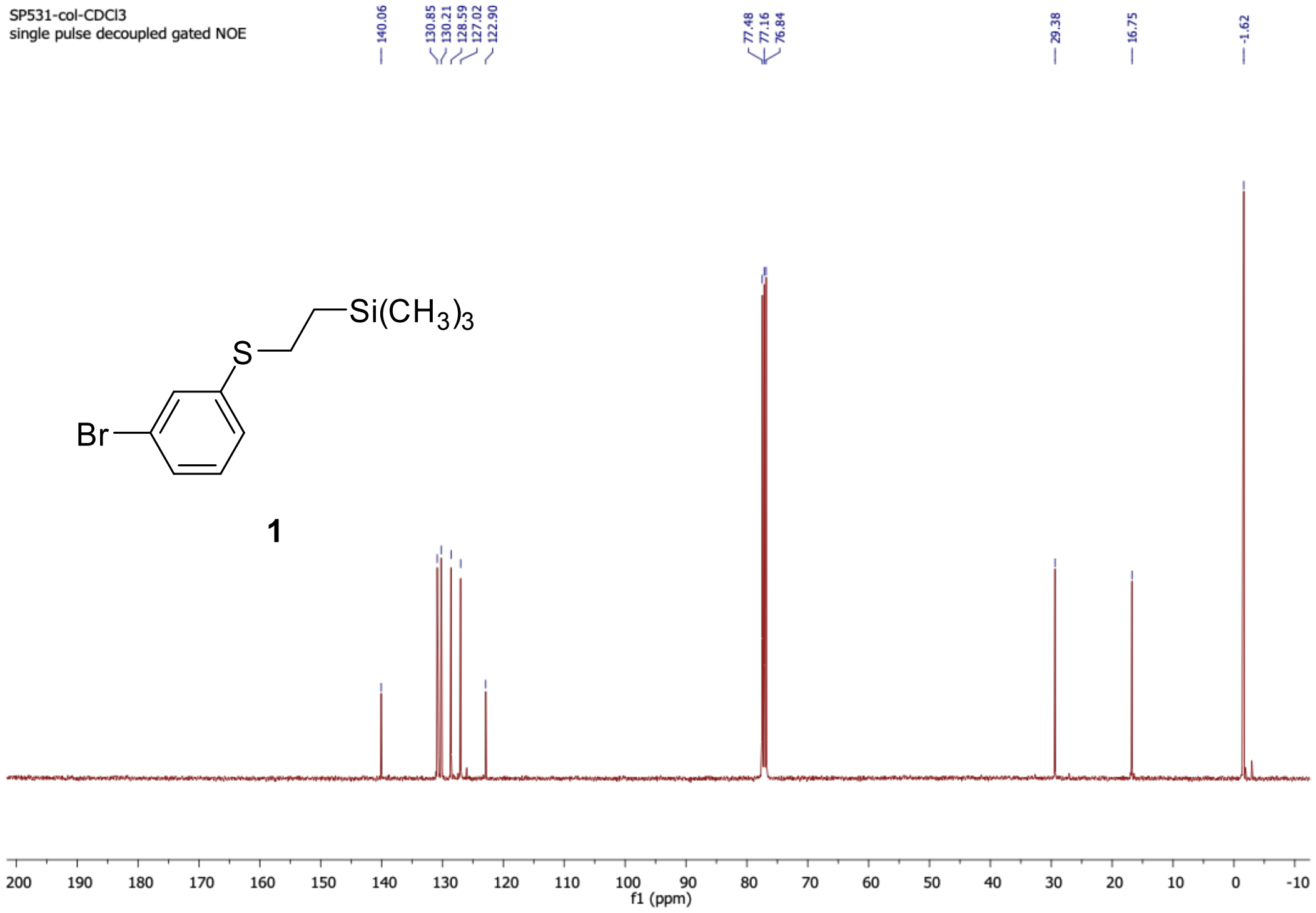

***Figure S2***. *[13]C{[1]H} NMR (101 MHz, $CDCl_3$) of compound* ***1***.

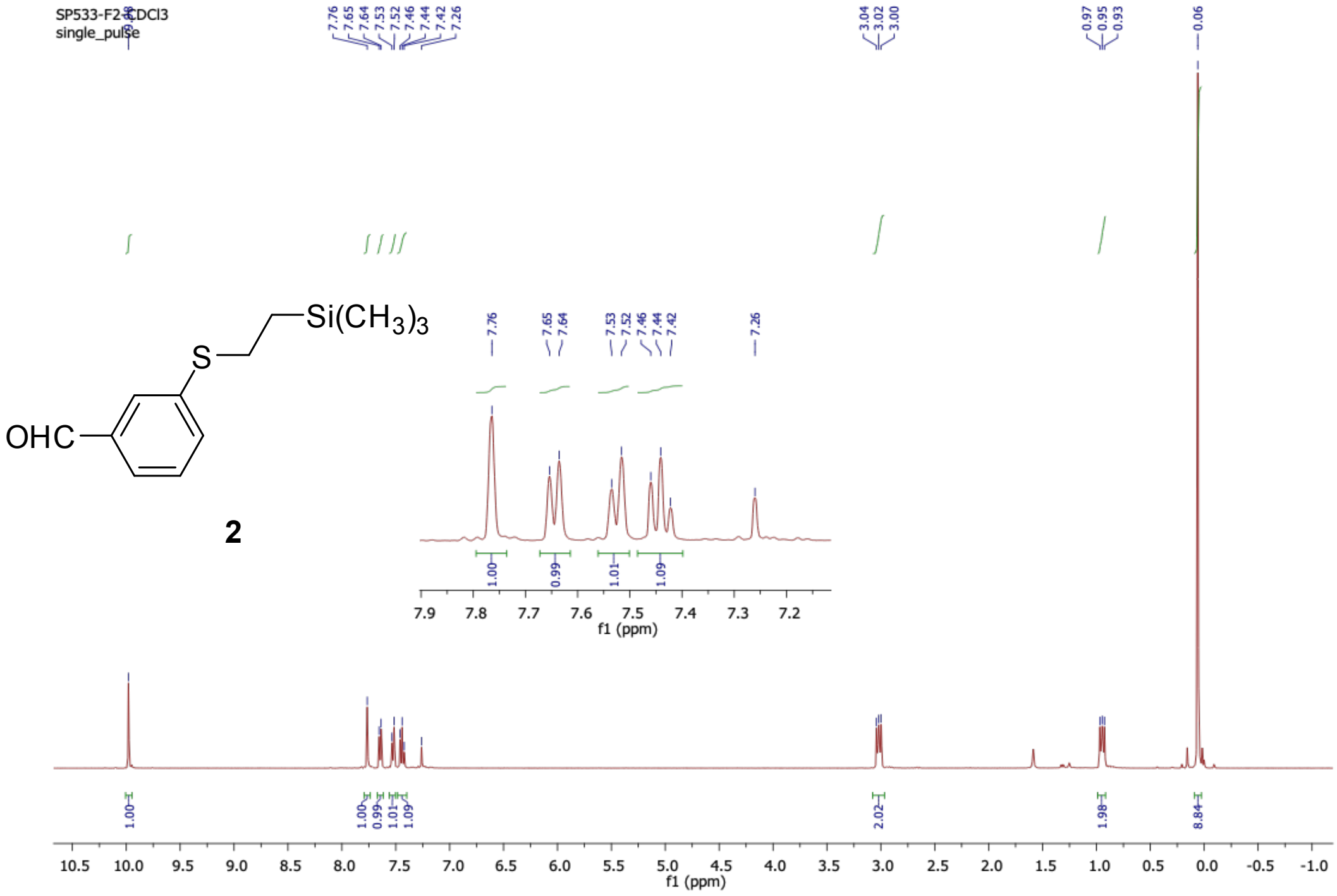

***Figure S3***. *[1]H NMR (400 MHz, $CDCl_3$) of compound* ***2***.

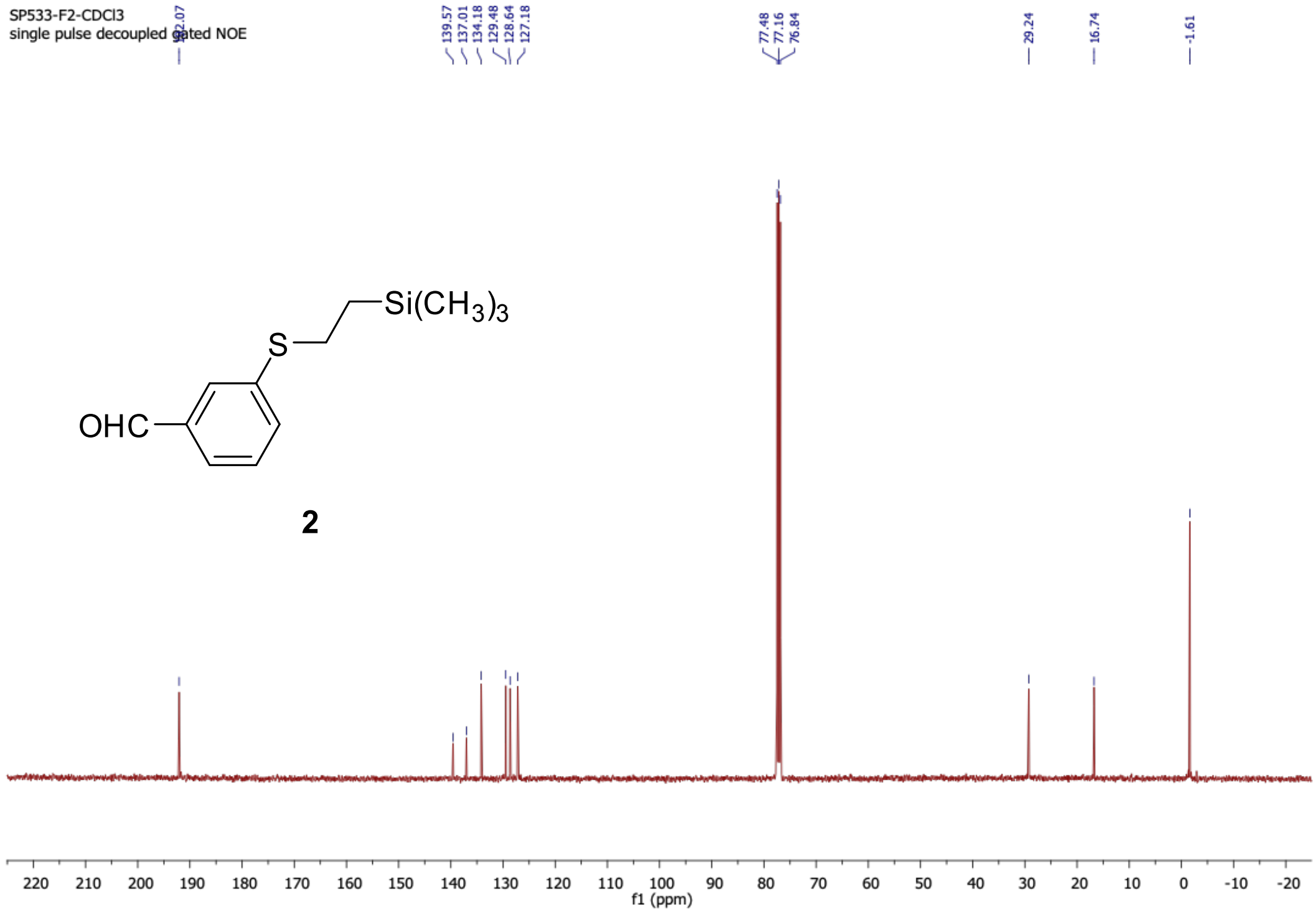

***Figure S4***. *$^{13}C\{^{1}H\}$ NMR (101 MHz, $CDCl_3$) of compound **2**.*

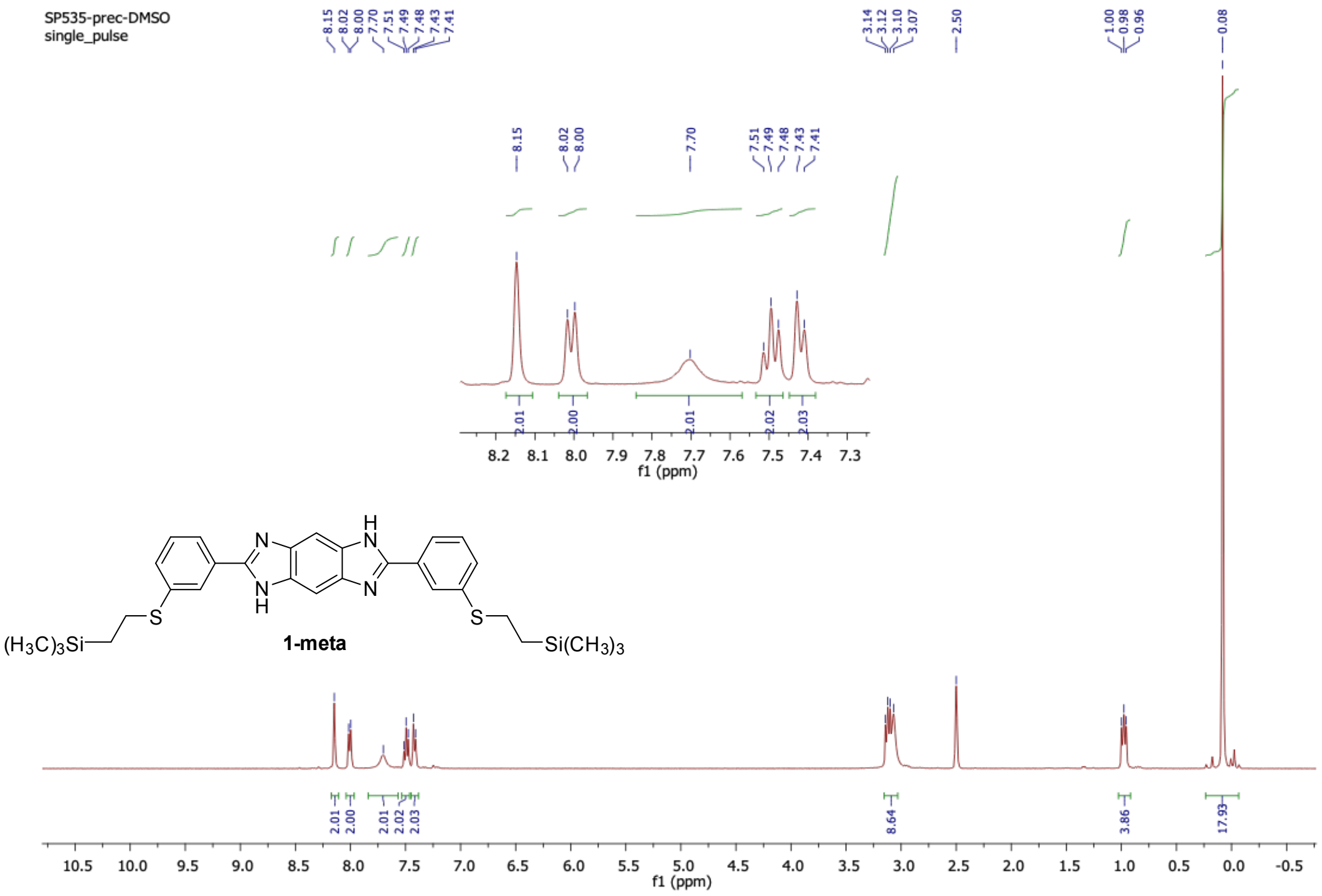

***Figure S5***. *$^{1}H$ NMR (400 MHz, DMSO-$d_6$) of **1-meta** recorded at 353 K.*

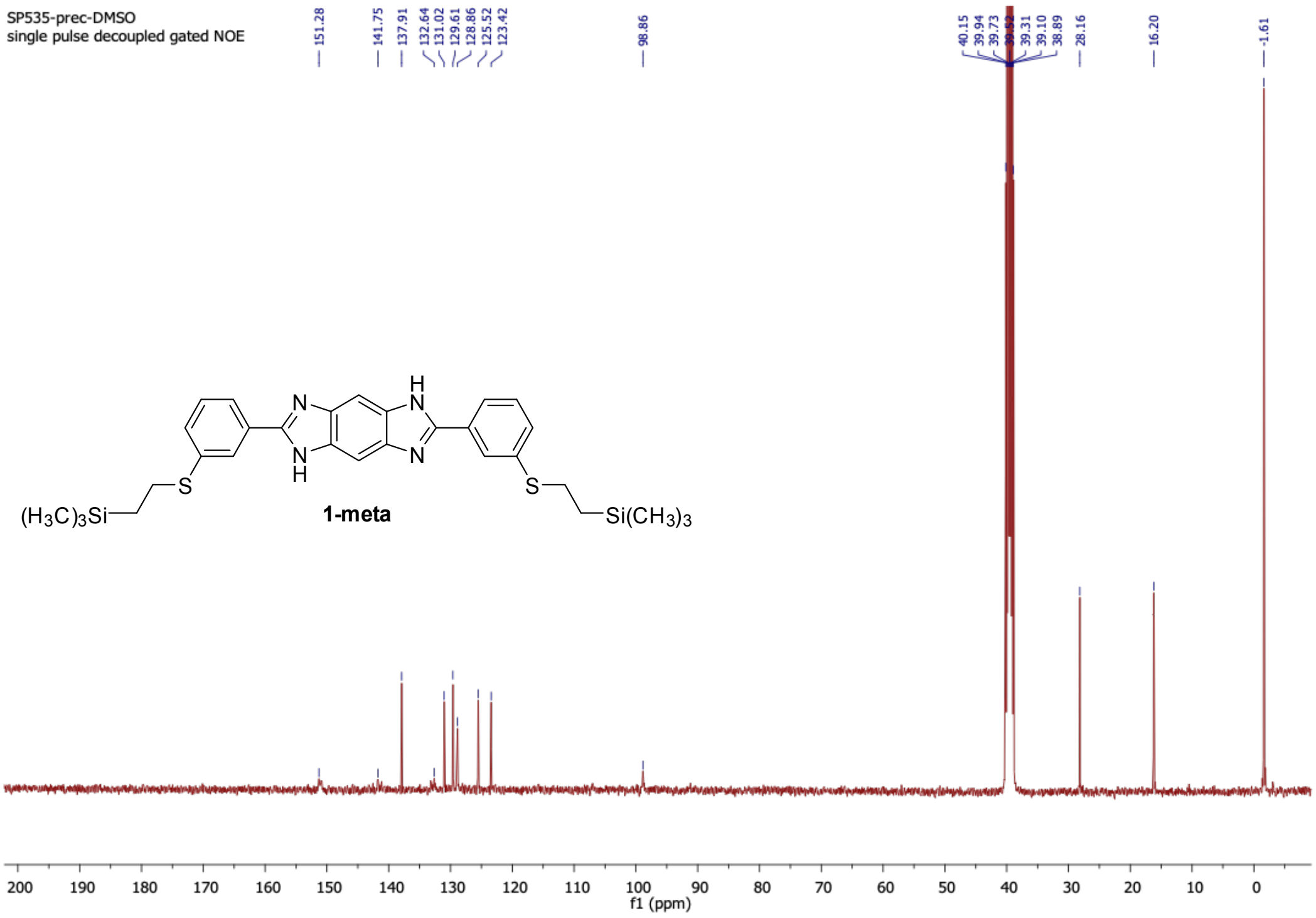

***Figure S6***. *$^{13}C\{^{1}H\}$ NMR (101 MHz, DMSO-$d_6$) of **1-meta**.*

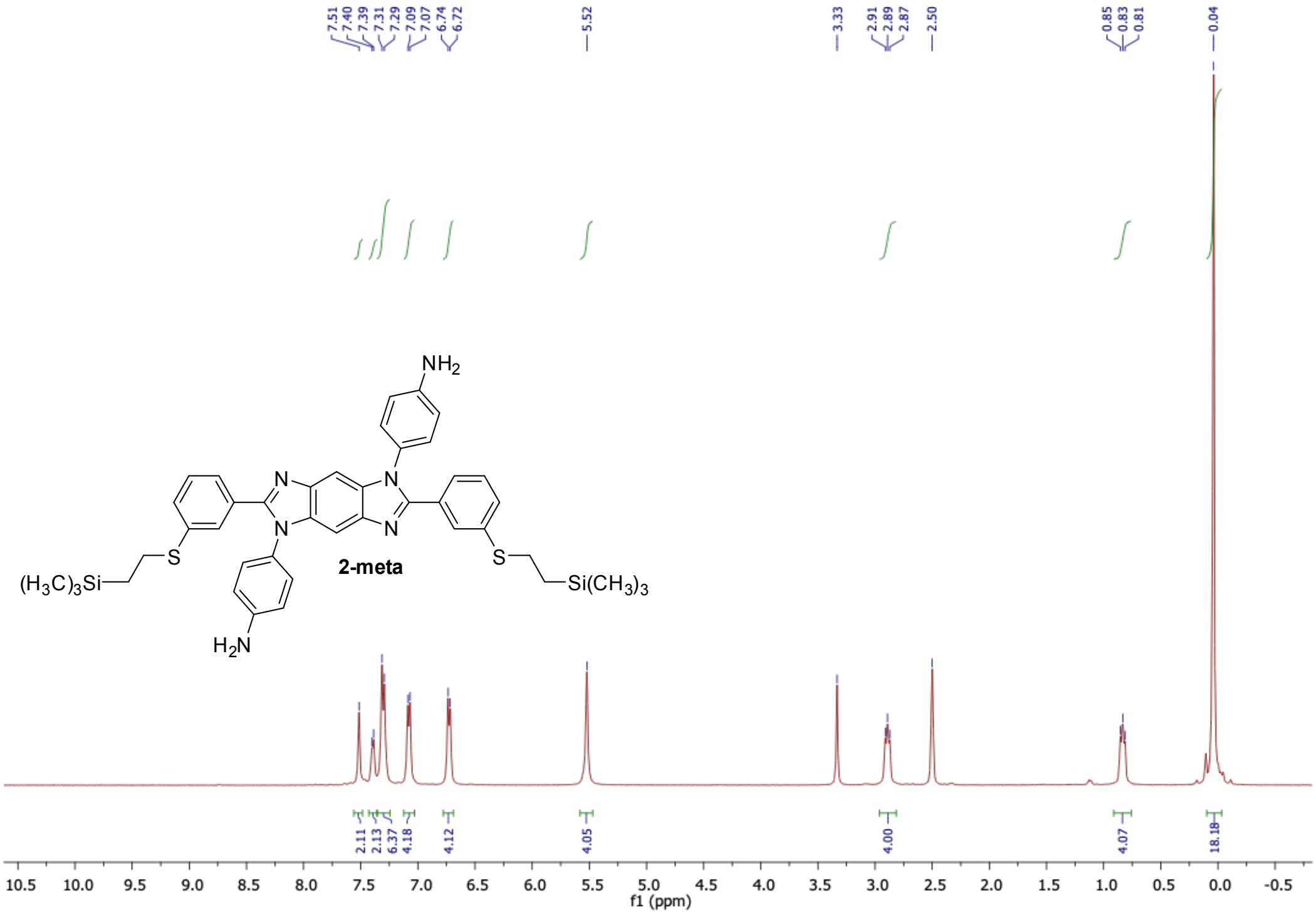

***Figure S7***. *$^{1}H$ NMR (400 MHz, DMSO-$d_6$) of **2-meta**.*

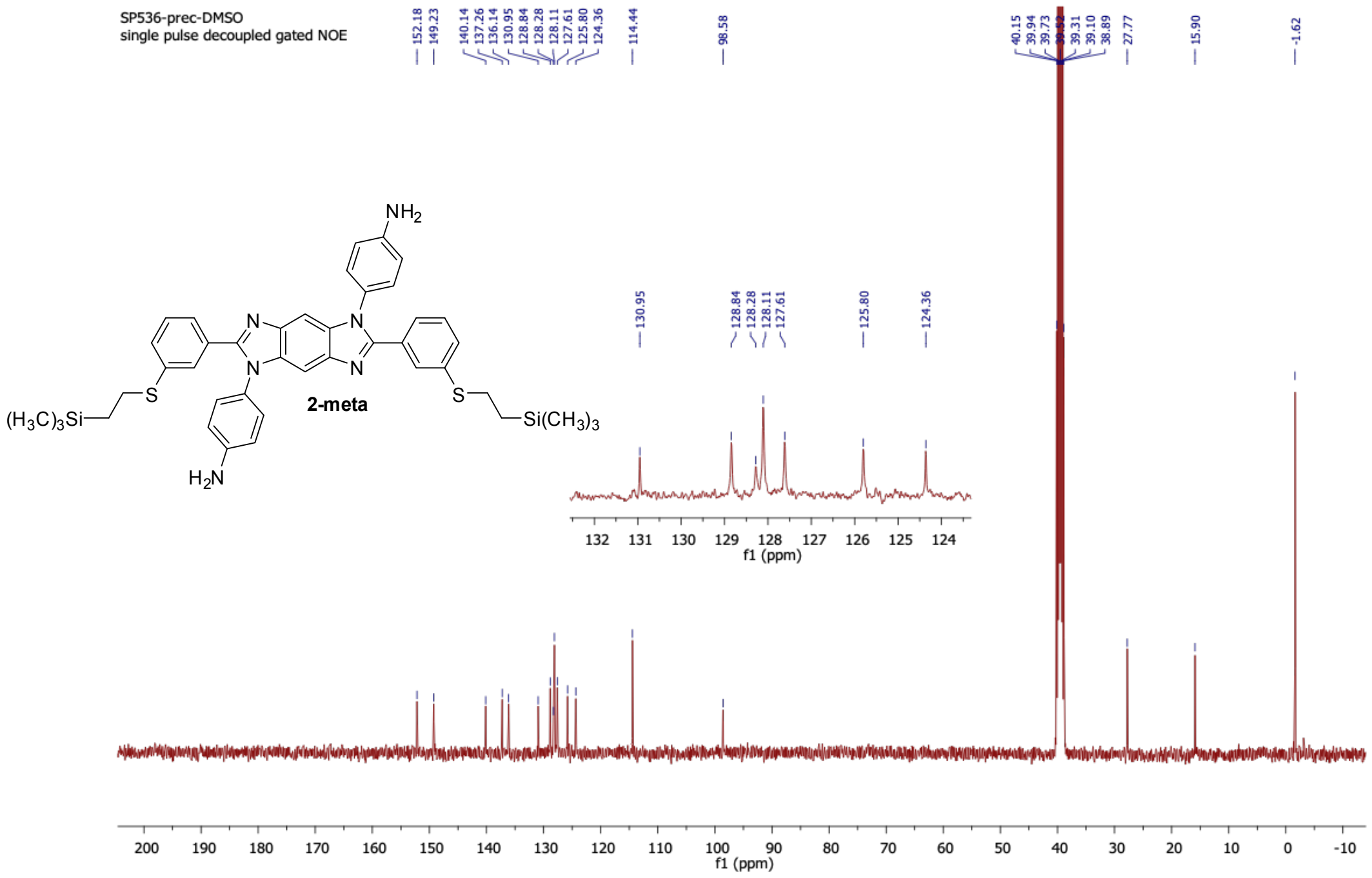

***Figure S8***. *$^{13}C\{^1H\}$ NMR (101 MHz, DMSO-d$_6$) of* ***2-meta***.

***Infrared spectroscopy.***

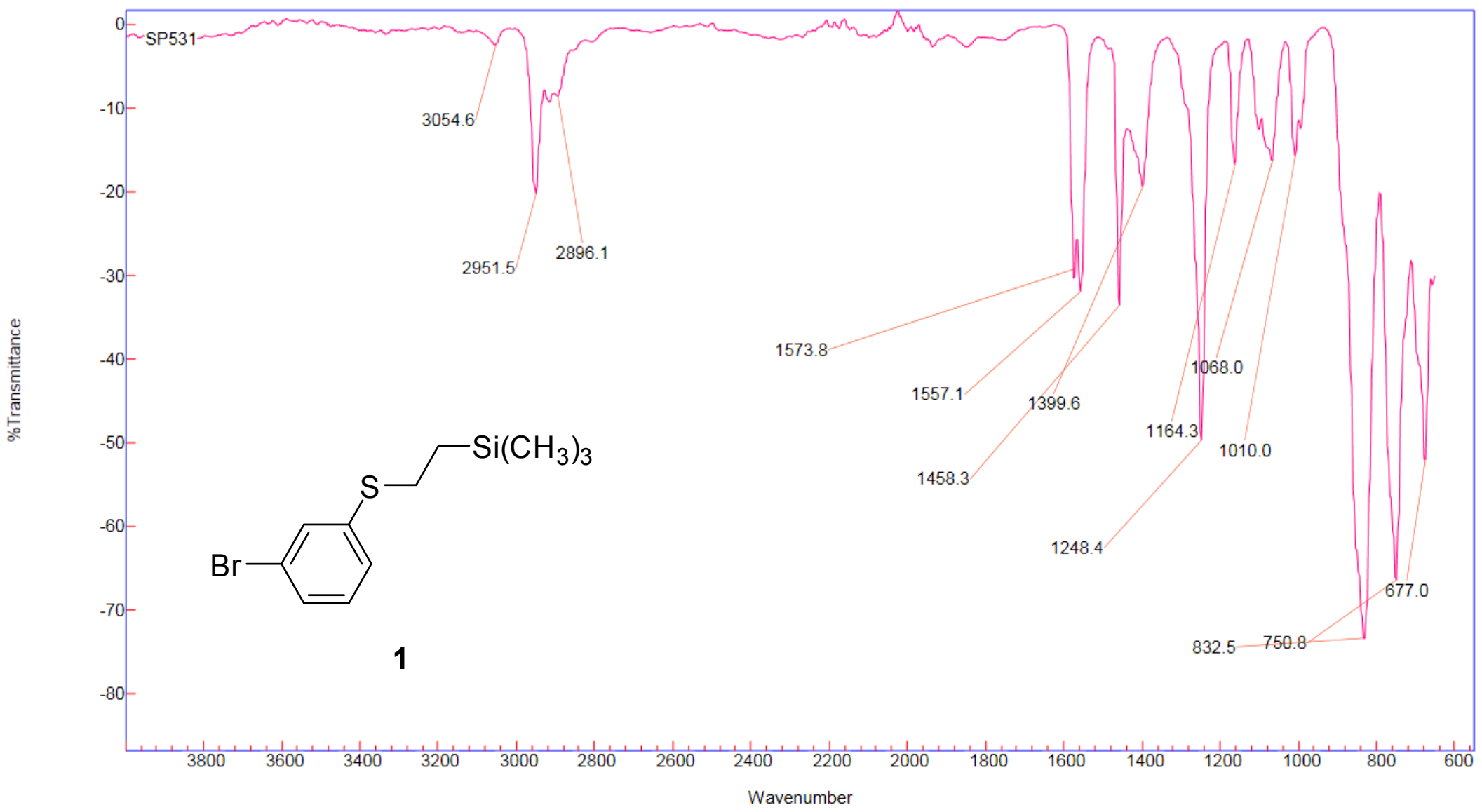

***Figure S9***. *Infrared spectrum (neat) of compound* ***1***.

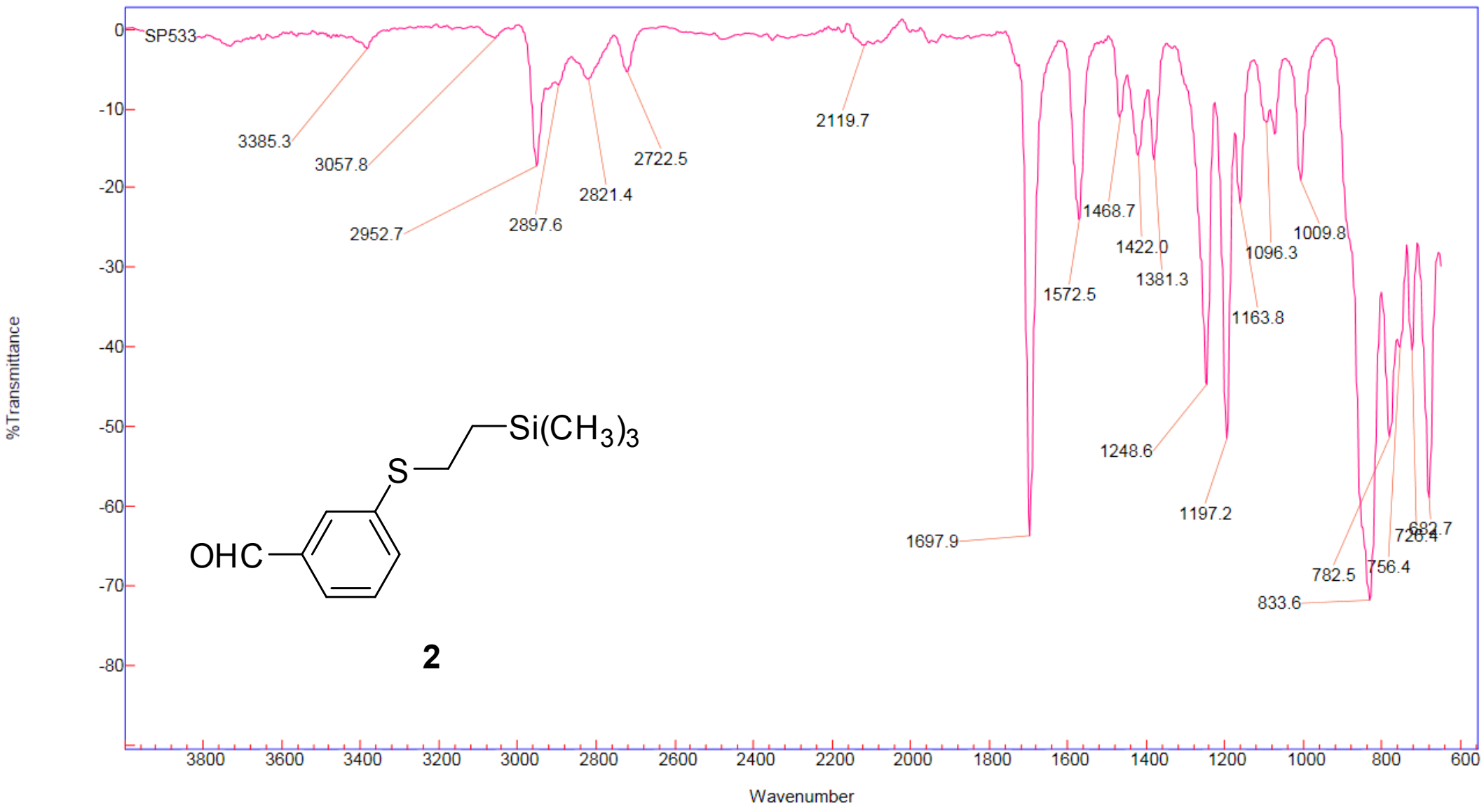

2

***Figure S10***. *Infrared spectrum (neat) of compound **2**.*

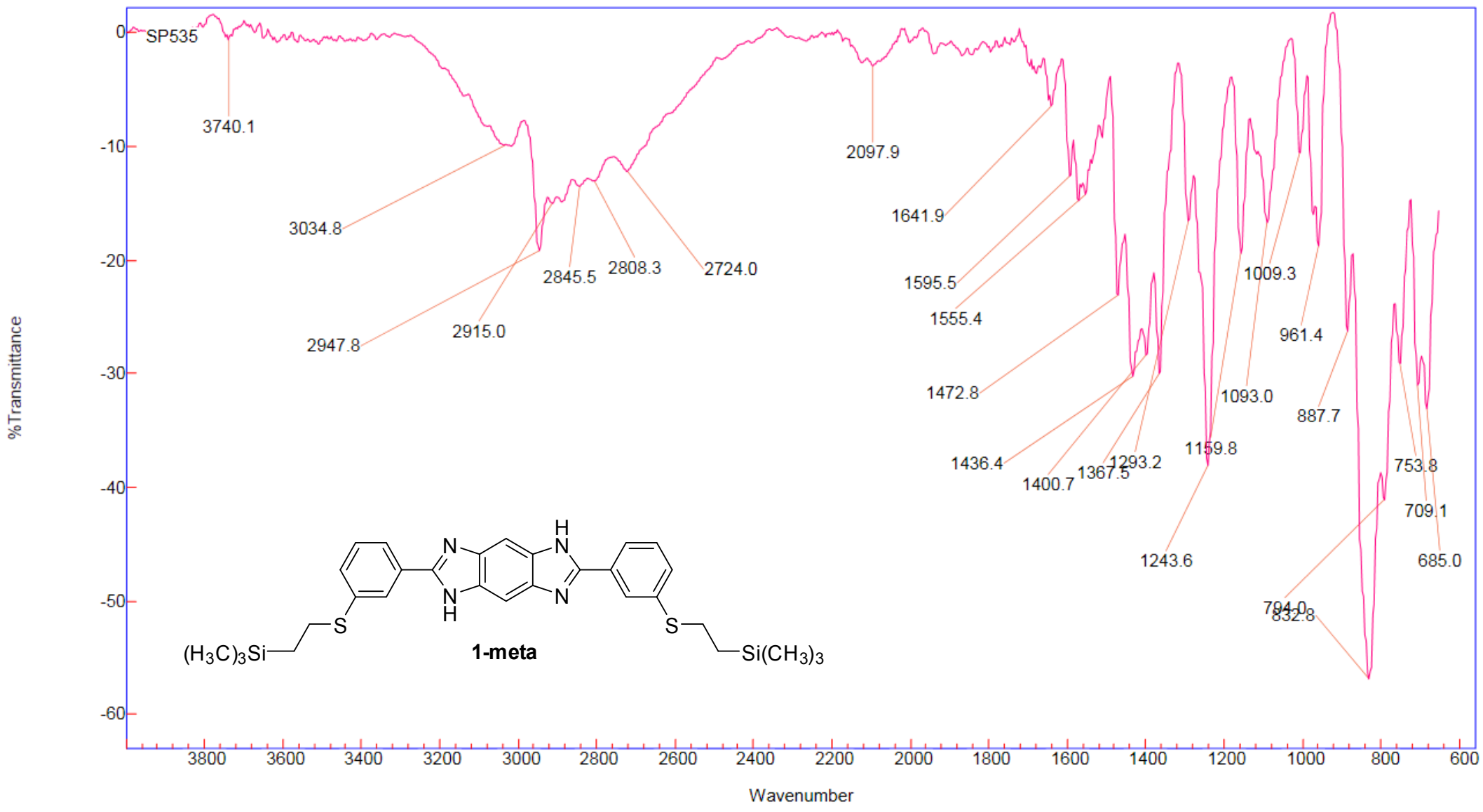

***Figure S11***. *Infrared spectrum (neat) of compound **1-meta**.*

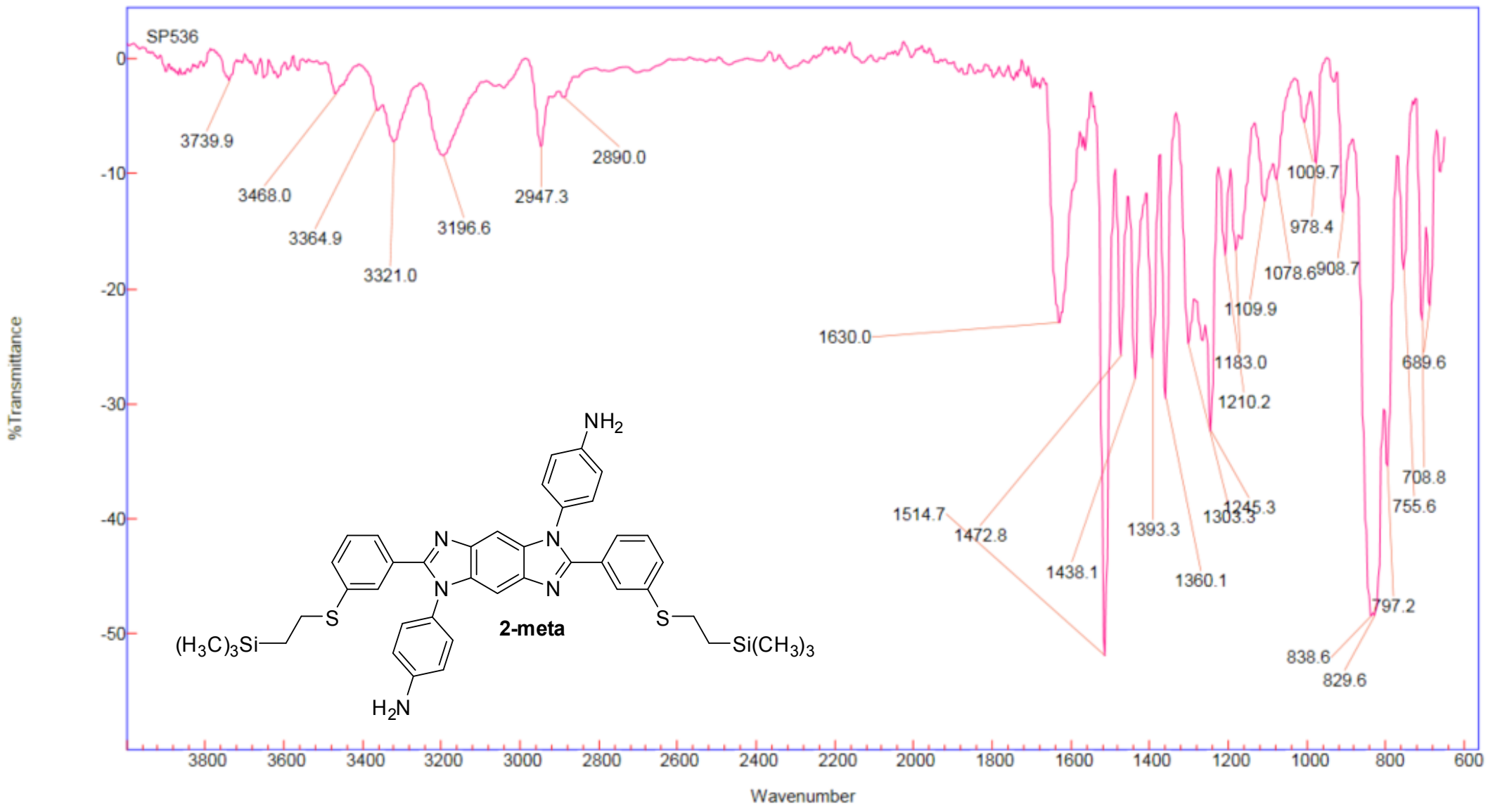

***Figure S12**. Infrared spectrum (neat) of compound **2-meta**.*

### ***High-resolution mass spectroscopy.***

The atomic mass of the synthesized compounds were checked by HRMS (Figs. S13-S16). The indicated m/z value corresponds to the isotopic cluster maximum (second isotopic peak).

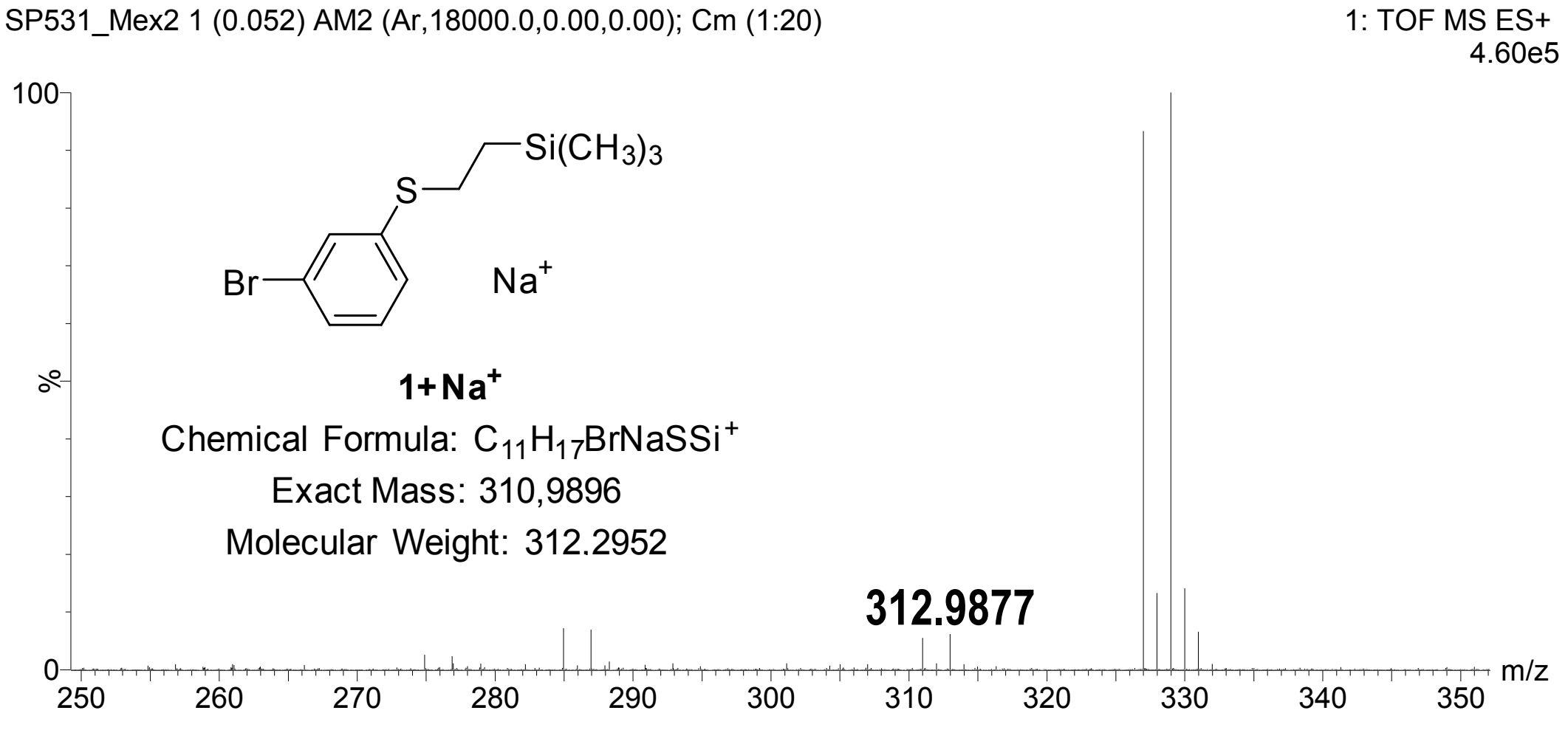

***Figure S13**. HRMS spectrum of compound 1.*

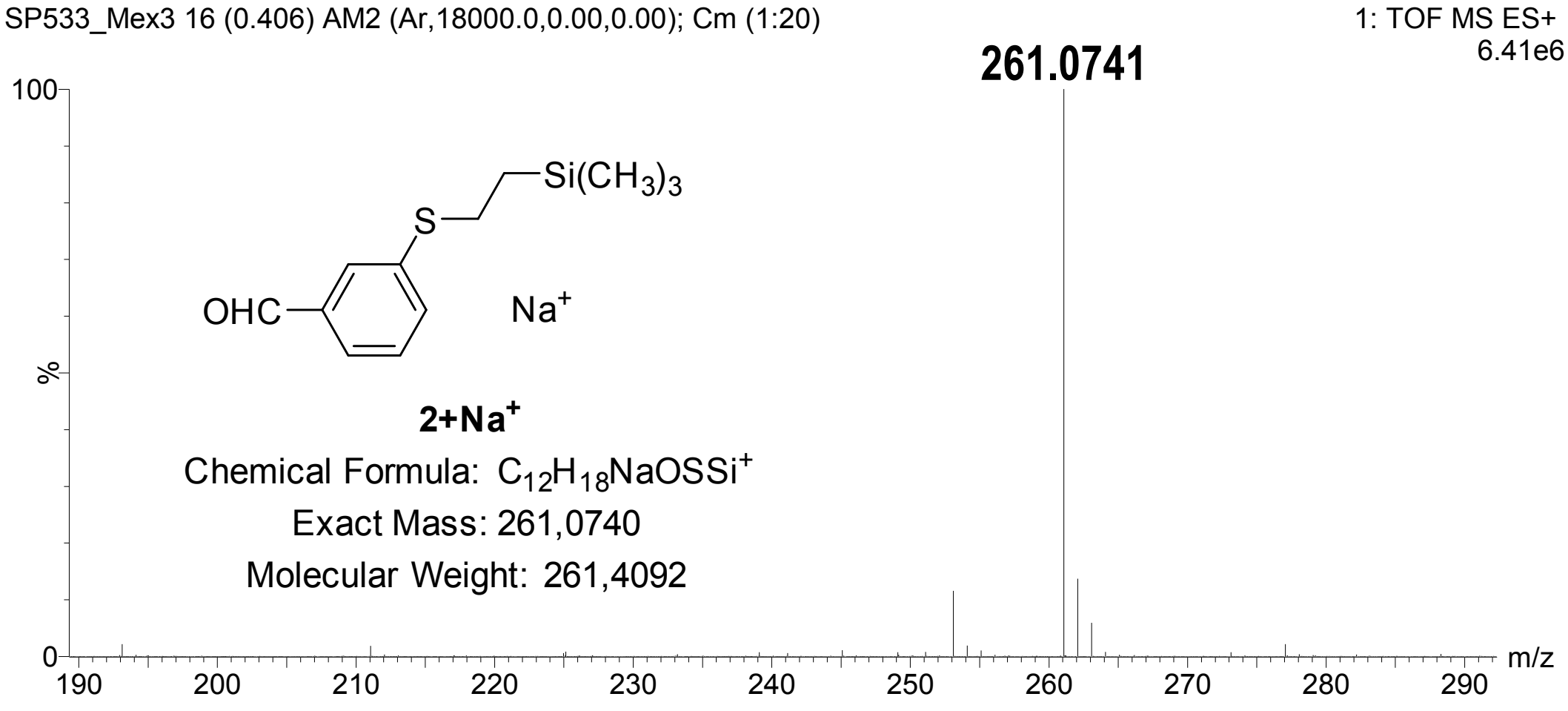

***Figure S14**. HRMS spectrum of compound **2**.*

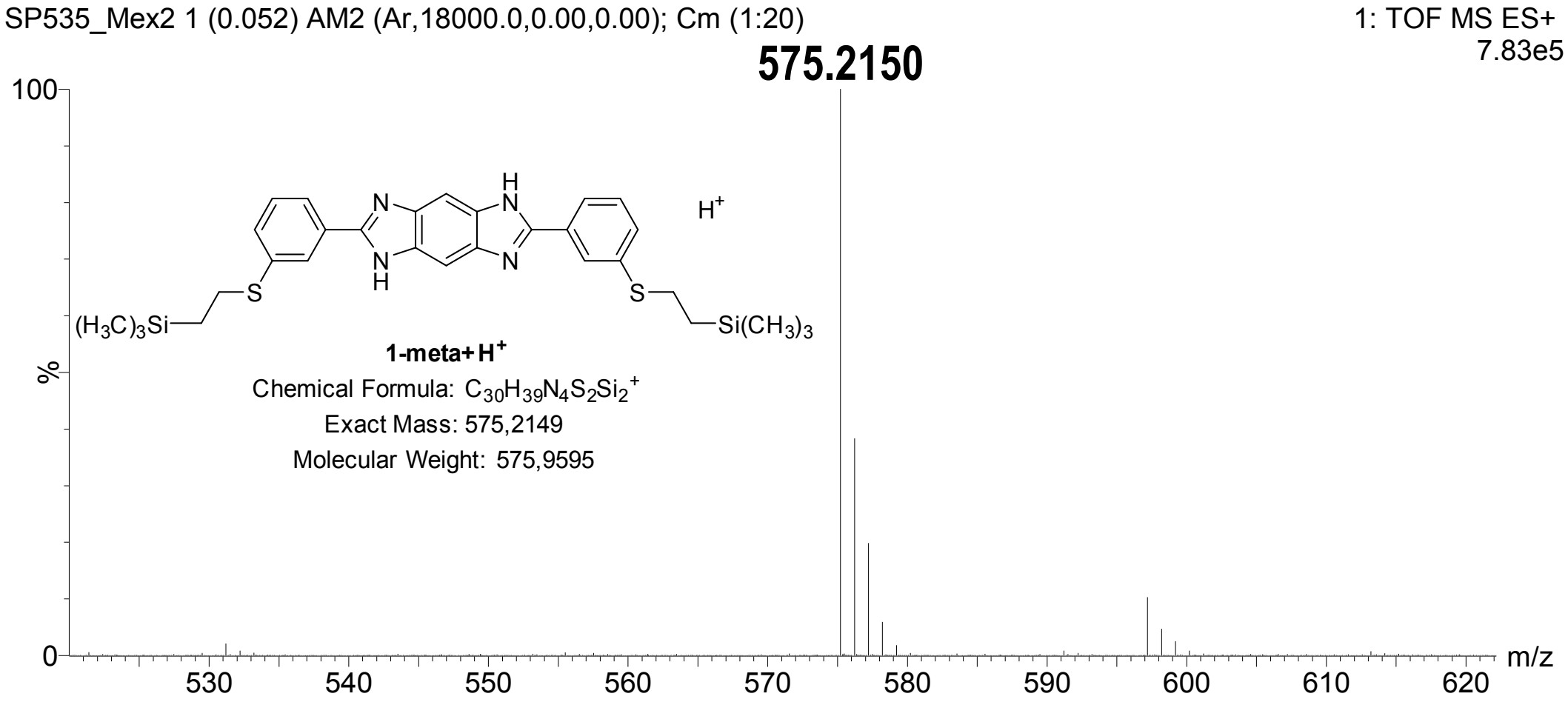

***Figure S15**. HRMS spectrum of compound **1-meta**.*

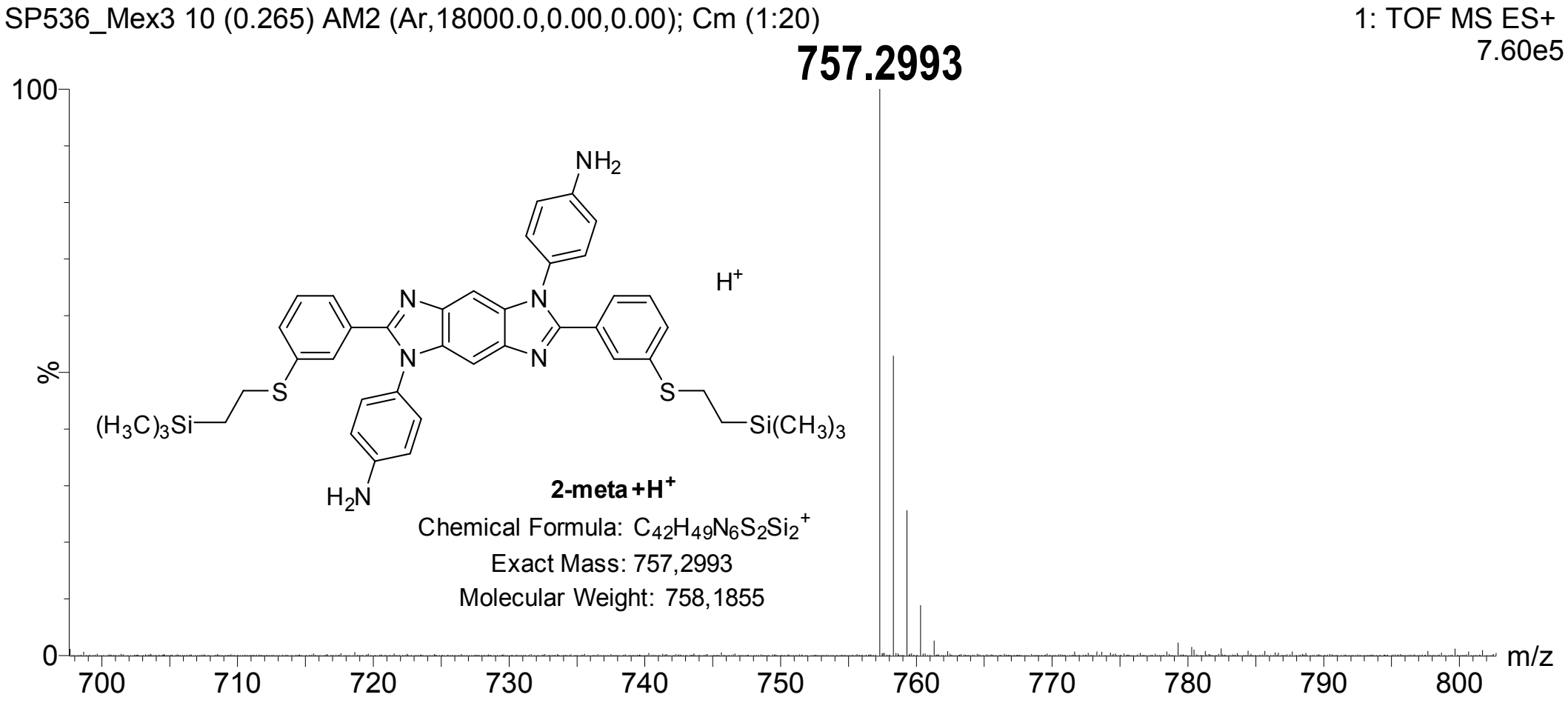

***Figure S16**. HRMS spectrum of compound **2-meta**.*

## II. Self-assembled monolayers on Au electrodes.

Ultraflat template-stripped gold surfaces ($^{TS}$Au), with rms roughness of ~0.4 nm were prepared according to methods already reported.[1-3] In brief, a 300−500 nm thick Au film was evaporated on a very flat silicon wafer covered by its native $SiO_2$ (rms roughness of ~0.4 nm), which was previously carefully cleaned by piranha solution (30 min in 7:3 H2SO4/$H_2O_2$ (v/v)); **Caution**: Piranha solution is a strong oxidizer and reacts exothermically with organics), rinsed with deionized (DI) water, and dried under a stream of nitrogen. Clean 10x10 mm pieces of glass slide (ultrasonicated in acetone for 5 min, ultrasonicated in 2-propanol for 5 min, and UV irradiated in ozone for 10 min) were glued to the evaporated Au film (UV-polymerizable glue, NOA61 from Epotecny), then mechanically peeled off providing the $^{TS}$Au film attached on the glass side (Au film is cut with a razor blade around the glass piece).

The self-assembled monolayers (SAMs) of **1-*meta*** and **2-*meta*** were prepared from 1 mM solution of the molecules unprotected by $NBu_4F$ in DMSO:ethanol (50:50). The $^{TS}$Au substrates were immersed for 3 days (in a glove box and in the dark).

## III. Ellipsometry.

We recorded spectroscopic ellipsometry data (on ca. 1 $cm^2$ samples) in the visible range using a UVISEL (Horiba Jobin Yvon) spectroscopic ellipsometer equipped with DeltaPsi 2 data analysis software. The system acquired a spectrum ranging from 2 to 4.5 eV (corresponding to 300–750 nm) with intervals of 0.1 eV (or 15 nm). The data were taken at an angle of incidence of 70°, and the compensator was set at 45°. We fit the data by a regression analysis to a film-on-substrate model as described by their thickness and their complex refractive indexes. First, a background for the substrate before monolayer deposition was recorded. We acquired three reference spectra at three different places of the surface spaced of a few mm. Secondly, after the monolayer deposition, we acquired once again three spectra at three different places of the surface and we used a 2-layer model (substrate/SAM) to fit the measured data and to determine the SAM thickness. We employed the previously measured optical properties of the substrate (background), and we fixed the refractive index of the monolayer at 1.50.[4] SAMs on Au substrate as thin as 0.5 nm can be measured by ellipsometry.[5] The three spectra measured on the sample were fitted separately using each of the three reference spectra, giving nine values for the SAM thickness. We calculated the mean value from these nine thickness values and the thickness incertitude corresponding to the standard deviation. Overall, the accuracy of the SAM thickness measurements is estimated at ± 2 Å.[6] Figure S17 shows the geometric optimized length (L) of the molecules (in gas phase, MM2 level, Chem3D). The measured thicknesses ($t_{SAM}$) are smaller than the length (-S to S- atoms). We estimate an average tilt angle Φ of the long axis of the molecules with respect to the Au surface normal by $\cos \Phi = t_{SAM}/L$. The area per molecule is estimated considering that the molecule is modeled by a cylinder, a tiny rod, of length L and a base diameter equal to the width (W) of the molecule (Fig. S17).[4, 5] The area per molecule is the projected base area on the substrate surface as defined by its projected diameter on the substrate surface, W/cos Φ. The values are summarized in Table S1. From this simple estimate, we infer that the *meta*-connected SAMs are less packed than the *para*-connected SAMs.

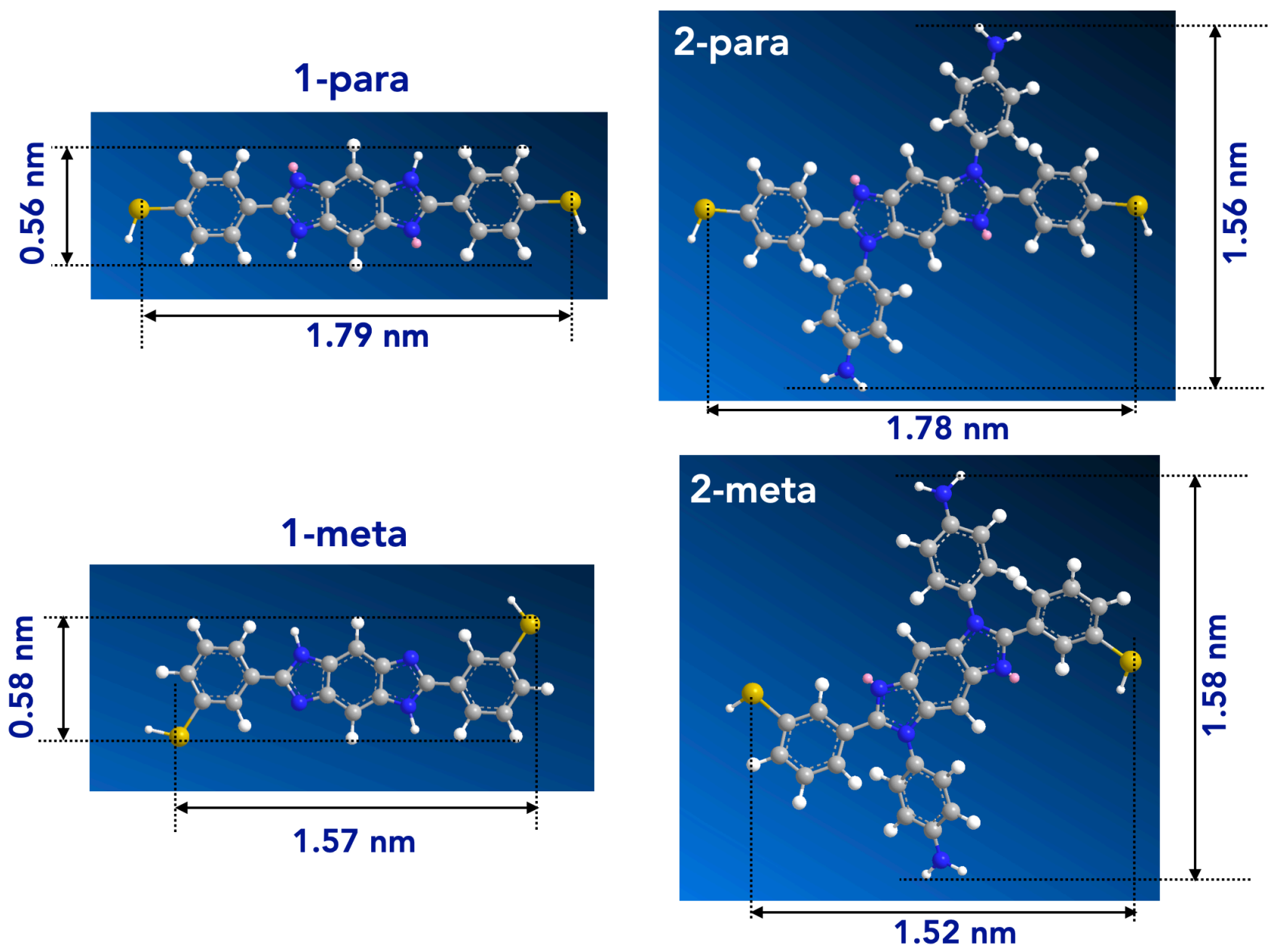

***Figure S17***. *Geometry optimization length (S-to-S atoms) and width in the gas phase of the four molecules (Chem 3D, MM2 level).*

| | 1-para | 2-para | 1-meta | 2-meta |
|---|---|---|---|---|
| Measured thickness (nm) | 1.85 | 1.4 | 0.6 | 0.6 |
| Theoretical length : S-to-S (nm) | 1.79 | 1.78 | 1.57 | 1.52 |
| Tilt angle (°) | 0 | 38 | 67 | 67 |
| Theoretical diameter (nm) | 0.56 | 1.56 | 0.58 | 1.58 |
| Projected diameter (nm) | 0.56 | 1.98 | 1.48 | 4.0 |
| Area per molecule (nm$^2$) | 0.25 | 3.08 | 1.73 | 12.58 |

***Table S1***. *Measured thickness ($t_{SAM}$) of the SAMs and comparison with the molecule length (L). The S-to-S length L is taken from the geometry optimization in the gas phase shown in Fig. S17. We also give the tilt angle Φ from the surface normal (cos Φ = $t_{SAM}$/L), the molecule width (W) considered as the cylinder model diameter, the projected diameter on the surface (W/cos Φ), and the corresponding area per molecule occupied on the surface.*

## IV. Null-point SThM.

### *1. General procedure.*

SThM[7, 8] were carried out with a Bruker ICON instrument equipped with the Anasys SThM module and in an air-conditioned laboratory (22.5°C and a relative humidity of 35-40%). We used Kelvin NanoTechnology (KNT) SThM probes with a Pd thin film resistor in the probe tip as the heating element (VITA-DM-GLA-1). The SThM tip is inserted into a Wheatstone bridge; the heat flux through the tip is controlled by the DC voltage applied on the Wheatstone bridge ($V_{WB}$, typically 0.6-1.1 V). The tip temperature, $T_{tip}$, is obtained by measuring the output voltage of the Wheatstone bridge, knowing the transfer function of the bridge, the gain of the voltage amplifier and the calibrated linear relationship between the tip resistance and the tip temperature.

The null-point SThM[9] was used at selected points on the SAMs. We define a 5x5 grid, each point spaced by 10 nm. At each point of the grid, in the z-trace mode (approach and retract), we recorded the tip temperature versus distance curve ($T_{tip}$-$z$). At the transition from a non-contact (NC, tip very near the surface) to a contact (C, tip on the surface) situation, we observe a temperature jump, $T_{NC}$ - $T_C$ (Fig. 5, main text), which is used to determine the sample thermal conductivity according to the protocol described in Ref. 9. The temperature jump is measured from the approach trace only (to avoid any artifact due to well-known adhesion hysteresis of the retract curve) and averaged over the 25 recorded $T_{tip}$-$z$ traces. This differential method is suitable for removing the parasitic contributions (air conduction, etc…): at the contact (C) both the sample and parasitic thermal contributions govern the tip temperature, whereas, just before physical tip contact (NC), only the parasitic thermal contributions are involved. The plot of the temperature jump, $T_{NC}$ - $T_C$, versus the sample temperature at contact $T_C$ is linear and its slope is inversely proportional to the thermal conductivity (Figs. 5, S19, S20). The tip-sample temperature $T_C$ increases with the supply voltage of the Wheatstone bridge $V_{DC}$ (typically from 0.6 to 1.1 V). The SAM/Au thermal conductance, $G_{th}$(SAM/Au), is calculated with Eq. S1:[9]

$$T_C - T_{amb} = \left( \alpha \frac{4r_{th}}{G_{th}(SAM/Au)} + \beta \right) (T_{NC} - T_C) \qquad \text{(S1)}$$

where $T_{amb}$ is the room temperature (22.5°C in our air-conditioned laboratory) and $r_{th}$ is the thermal contact radius of the SThM tip ≈ 20 nm (*vide infra*). The calibration parameters, α and β, depend on the peculiar tip and equipment. They were systematically measured before all the measurements. The same $T_C$ vs. $T_{NC}$-$T_C$ measurements were done on two materials with well-known thermal conductivity (or equivalently the thermal conductance, $G_{th}=4r_{th}\kappa$): a glass slide (κ = 1.3 W m$^{1}$ K$^{-1}$) and a low-doped silicon wafer with its native oxide (κ = 150 W m$^{-1}$ K$^{-1}$), Fig. S18. A new calibration was done for each new sample and new tips to cope with slight changes in the instrument parameters (*e.g.,* wear and tear of the tip, shift of loading force). Figure S18 shows a typical calibration curve. From a linear fit of the data (Fig. S18), we get $\alpha$ = 14.96 W m$^{-1}$ K$^{-1}$ and $\beta$ = 7.72 K/K in this typical example (calibration for the measurements on sample **1-*meta*** protonated).

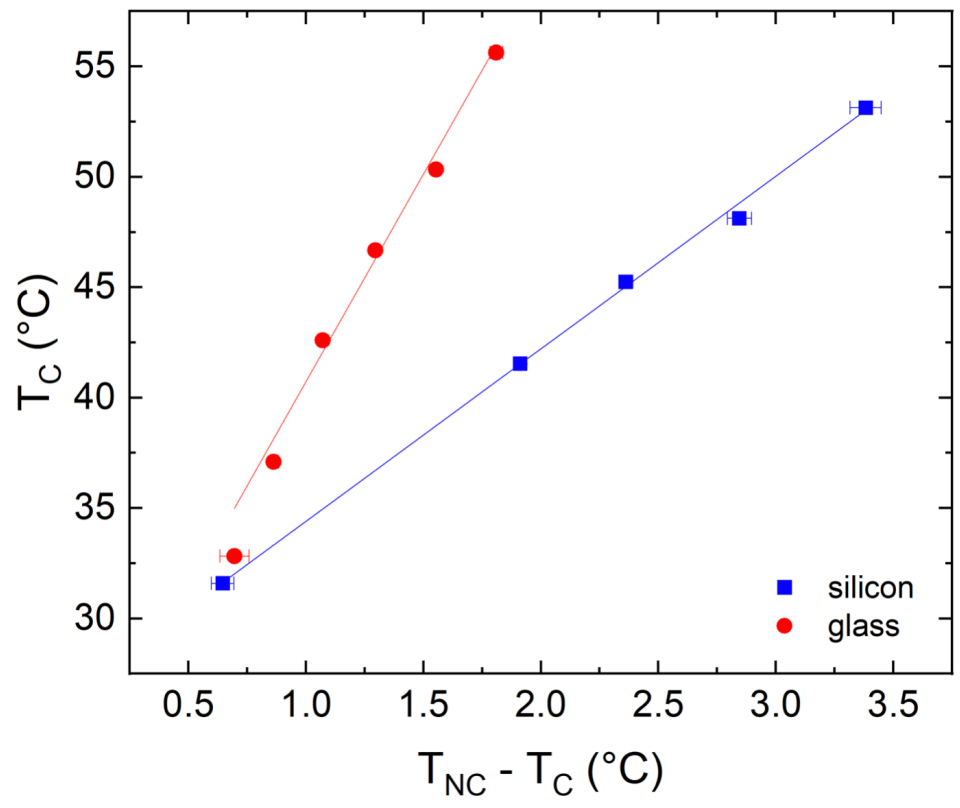

***Figure S18****. Typical calibration curve before the measurements on the* ***1-meta*** *protonated MJ.*

## *2. Correction with the Dryden model.*

The SAMs are deposited on a high thermal conducting Au substrate and the measured "effective" conductance $G_{th}$(SAM/Au) contains a contribution from the substrate. The Dryden model[10] allows calculating the constriction thermal conductance $G_{th}$ (or equivalently the thermal conductivity $\kappa$, $G_{th}=4r_{th}\kappa$) when a small thermal spot is contacting a thin layer coating on a substrate. We used this model to calculate the thermal conducta nce $G_{th}$(SAM) of a very thin film (here the SAM) of thickness $t_{SAM}$ deposited on a semi-infinite substrate (here the thick underlying Au electrode) with a thermal conductance $G_{th}$(Au) = 25.4 µW/K (with $\kappa_{Au}$ = 318 W m$^{-1}$ K$^{-1}$ and $r_{th}$ ≈ 20 nm, *vide infra*). In the case $t_{SAM}/r_{th} < 2$ the model reads

$$\frac{1}{G_{th}(SAM/Au)} = \frac{1}{G_{th}(Au)} + \frac{4}{\pi G_{th}(SAM)}\left(\frac{t_{SAM}}{r_{th}}\right)\left(1 - \left(\frac{G_{th}(SAM)}{G_{th}(Au)}\right)^2\right) \quad \text{(S2).}$$

Solving this equation for all the measured $G_{th}$(SAM/Au) allows determining the SAM thermal conductance $G_{th}$(SAM).

### 3. Thermal contact area.

The thermal contact radius (at the tip/SAM interface) is calculated following the approach reported in Ref. 11 taken into account the mechanical tip radius $r_{tip}$ and the size of the water meniscus at the tip/surface interface. The thermal radius of the thermal contact is estimated by[12]

$$r_{th} = 2.08\sqrt{\frac{-r_{tip}\cos\theta}{\ln\varphi}} \quad \text{(S3)}$$

with $r_{tip}$ = 100 nm (data from Bruker), the relative humidity $\varphi$ = 0.35-0.4 (in an air-conditioned laboratory, values checked during the measurements) and the contact angle of the concave meniscus between the tip and the surface $\theta \approx 30°$ as measured for π-conjugated molecular crystals in Ref. 13. We get $r_{th} \approx 20.2$ nm. The water meniscus contact angle depends on the surface energy of the sample, and thus should, in principle, depend on the chemical nature and molecule organization of the SAM and also likely on the protonation/deprotonation state of the molecules. However, we cannot perform water contact angle measurements inside the nanometer size tip/SAM interface, and we consider the same literature value of 30° in all cases. We can estimate a confidence interval by assuming that the protonation modifies the surface energy (making the SAM surface more hydrophilic) and that the water contact angle Θ decreases down to a few degrees. Let us consider a very weak value Θ ≈ 5° for a strong hydrophilic SAM surface,[4] we get $r_{th} \approx 21.7$ nm, *i.e.*, a relative variation of ≈ 8%. However, this error in determining $G_{th}$ (Eqs. S1 and S2) remains negligible compared to the error bars (Fig. 4-b, relative error ≈ 15-28 % from data in table S2) and the relative variations of $G_{th}$ upon protonation (≈ 47% and ≈ 17%, for ***1-meta*** and ***2-meta*** MJs, respectively, data from Table S2). Thus we have neglected this hydrophilic-related variation of $r_{th}$ to calculate $G_{th}$ of the protonated *meta*-connected MJs. This slightly underestimates (Eq. S1) the reported values of $G_{th}$. We note that, with this nominal value of the thermal contact radius, our SThM thermal conductances are in good agreement (relative errors ≲ 10 %) compared to other standard methods (photothermal radiometry, flash laser analysis, and micro-Raman), which are not sensitive to the tip-sample interface issues.[14] In our ambient air experiments, even in an air-conditioned environment with a controlled relative humidity, it is not possible to better control this contact angle. SThM in UHV would be further required to limit the influence of this parameter.[15]

## V. Conductive-AFM.

### 1. General procedure.

We measured the electron transport properties at the nanoscale by C-AFM (ICON, Bruker) at room temperature (in an air-conditioned laboratory: 22.5°C and a relative humidity of 35-40%) using a tip probe in platinum/iridium (PtIr), model SCM-PIC-V2 from Bruker. We used a "blind" mode to measure the current-voltage (*I-V*) curves and the current histograms: a square grid of 10×10 was defined with a pitch of 50 to 100 nm. At each point, one *I-V* curve is acquired, leading to the measurements of 100 traces per grid. This process was repeated several times at different places (randomly chosen) on the sample, and up to several thousands of *I-V* traces were used to construct the current-voltage histograms (as shown in Figs. 2 and 3, main text). The tip load force was set at ≈ 3-5 nN for all the *I-V* measurements, a lower value leading to too many contact instabilities during the *I-V* measurements.

### 2. C-AFM contact area.

As usually reported in literature[16-19] the contact radius, *a*, between the C-AFM tip and the SAM surface, and the SAM elastic deformation, *δ*, are estimated from a Hertzian model:[20]

$$a^2 = \left(\frac{3RF}{4E^*}\right)^{2/3} \tag{S4}$$

$$\delta = \left(\frac{9}{16R}\right)^{1/3}\left(\frac{F}{E^*}\right)^{2/3} \tag{S5}$$

with *F* the tip load force (≈ 5 nN), *R* the tip radius (20 nm) and *E** the reduced effective Young modulus defined as:

$$E^* = \left(\frac{1}{E^*_{SAM}} + \frac{1}{E^*_{tip}}\right)^{-1} = \left(\frac{1-\nu^2_{SAM}}{E_{SAM}} + \frac{1-\nu^2_{tip}}{E_{tip}}\right)^{-1} \tag{S6}$$

In this equation, $E_{SAM/tip}$ and $\nu_{SAM/tip}$ are the Young modulus and the Poisson ratio of the SAM and C-AFM tip, respectively. For the Pt/Ir (90%/10%) tip, we have $E_{tip}$ = 204 GPa and $\nu_{tip}$ = 0.37 using a rule of mixture with the known material data.[21] These parameters for the benzo-bis(imidazole) derivative SAMs are not known and, in general, they are not easily determined in such a monolayer material. Thus, we consider the value of an effective Young modulus of the SAM $E^*_{SAM}$ = 38 GPa as determined for the "model system" alkylthiol SAMs from a combined mechanic and electron transport study.[18] With these parameters, we estimated *a* ≈ 1.3 nm (contact area ≈ 6 nm²) and $\delta \approx 0.09$ nm. Considering the area per molecule on the surface, we can estimate the number, N, of molecules contacted by the tip. From the nominal lateral size of the molecule (Fig. S17, ≈ 0.5 nm and ≈ 1.5 nm for **1** and **2**, respectively, Table S1) and considering the molecule tilt angle (≈ 66°, in the *meta*-connected SAMs, see main text), we roughly estimated a molecular packing density of 1.7 nm²/molecule and 12.6 nm²/molecule for the **1-*meta*** molecule and **2-*meta*** SAMs respectively (Table S1). From these areas per molecule, we infer that about 1 and 4 molecules are measured in the SAM/PtIr junction for the **2-*meta*** and **1-*meta*** SAMs, respectively.

### 3. *Data analysis.*

Before constructing the current-voltage datasets shown in Figs. 2 and 3 (main text) and analyzing the *I-V* curves with the one energy-level model, the raw set of *I-V* data is scanned and some *I-V* curves were discarded from the analysis:

- At high current, the *I-V* traces that reached the saturating current during the voltage scan (the compliance level of the trans-impedance amplifier, typically $5\times10^{-8}$ A here (depending on the gain of the amplifier) and/or *I-V* traces displaying large and abrupt steps during the scan (contact instabilities).
- At low currents, the *I-V* traces that reached the sensitivity limit (almost flat *I-V* traces and noisy *I-V*s) and displayed random staircase behavior (due to the sensitivity limit - typically 2-3 pA depending on the used gain of the trans-impedance amplifier and the resolution of the ADC (analog-digital converter).

These specific cases can be encountered when the C-AFM tip is randomly located (see section V.1) on defects in the SAMs, such as pinholes (high currents) or molecular aggregates (low currents), for instance.

### 4. *Fits of the I-V curves with the analytical SEL model.*

All the I-V traces of the dataset of the molecular junctions were fitted with the single-energy level (SEL) model given by the following analytical expression:[22, 23]

$$I(V) = N\frac{8e}{h}\frac{\Gamma_1\Gamma_2}{\Gamma_1+\Gamma_2}\left[arctan\left(\frac{\varepsilon_H+\frac{\Gamma_1}{\Gamma_1+\Gamma_2}eV}{\Gamma_1+\Gamma_2}\right)-arctan\left(\frac{\varepsilon_H-\frac{\Gamma_2}{\Gamma_1+\Gamma_2}eV}{\Gamma_1+\Gamma_2}\right)\right] \quad (S7)$$

with $\varepsilon_H$ the energy of the HOMO involved in the transport (with respect to the Fermi energy of the electrodes), $\Gamma_1$ and $\Gamma_2$ the electronic coupling energy between the MO and the electron clouds in the two electrodes, *e* the elementary electron charge, *h* the Planck constant. This model neglects the intermolecular interactions[24-26] between neighboring molecules in the SAMs, and assumes independent molecules conducting in parallel, with $N$ the number of molecules contributing to the ET in the molecular junction. It was experimentally demonstrated that these interactions are negligible in less packed SAMs.[26] In the present case of meta-connected SAMs, which have a low packing density and a weak number of molecules in the MJs ($N$ = 1 and 4 for the **2-*meta*** and **1-*meta*** MJs, respectively, *vide supra*), this assumption is valid. Strictly speaking, this model is valid at 0 K, since the Fermi function of the electrodes is not taken into account. However, it was shown that it can be used to describe data measured at RT for voltages below the transition between the off-resonant and resonant transport conditions (i.e., for $|eV| < 2\varepsilon_H$ in the symmetric case $\Gamma_1 = \Gamma_2$) at which the broadening of the Fermi function modifies the *I-V* shape.[27-29] Since the applied voltage is limited to -0.5/0.5 V in all the I-V measurements (Figs. 2 and 3, main text), this condition is fulfilled.

## VI. Additional data and figures.

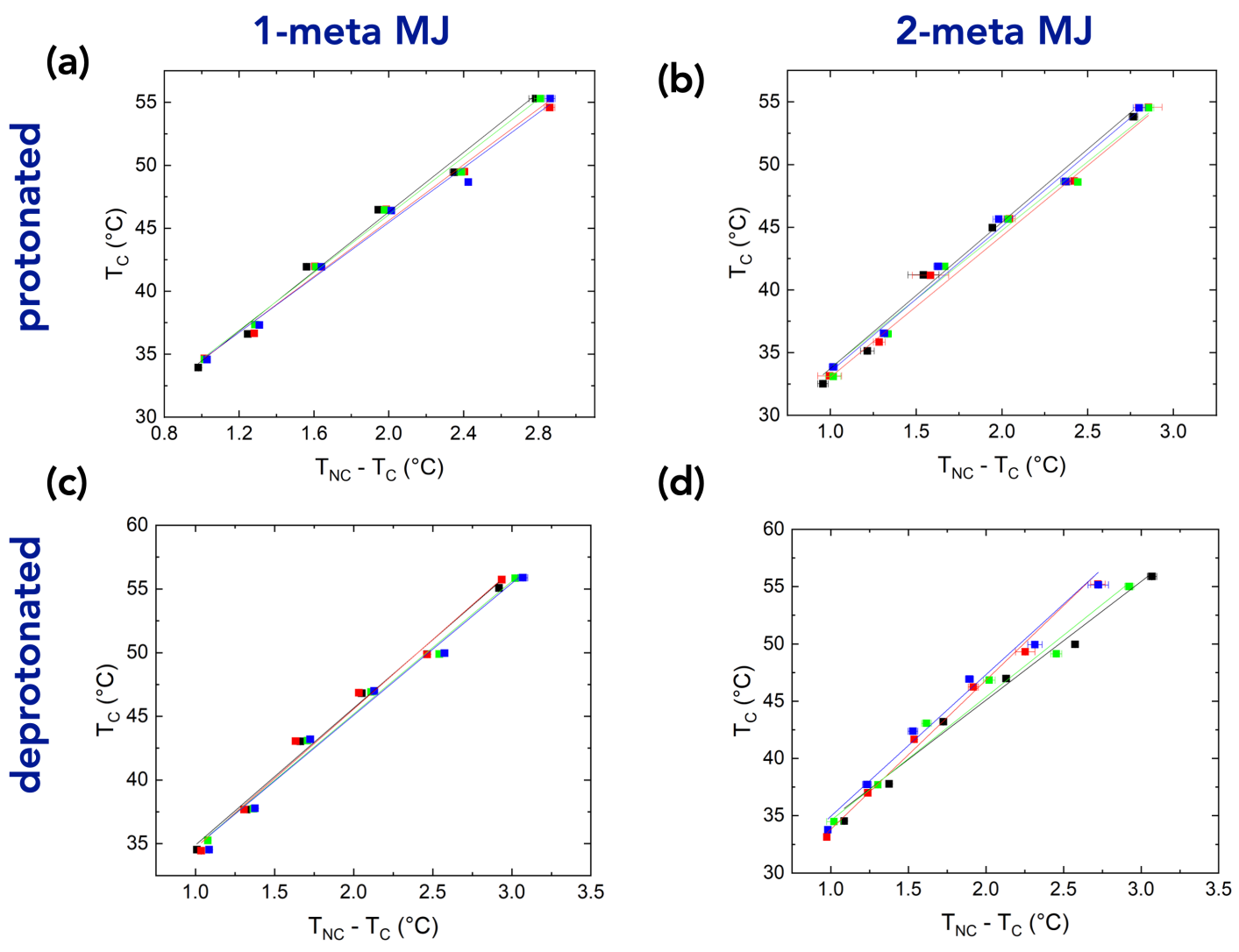

***Figure S19**. $T_C$ vs. $T_{NC}$-$T_C$ plots at 4 different locations on the SAMs for the **1-meta** and **2-meta** SAMs, in the protonated and deprotonated states. The lines are linear fits to calculate the sample thermal conductance $G_{th}$(SAM/Au).*

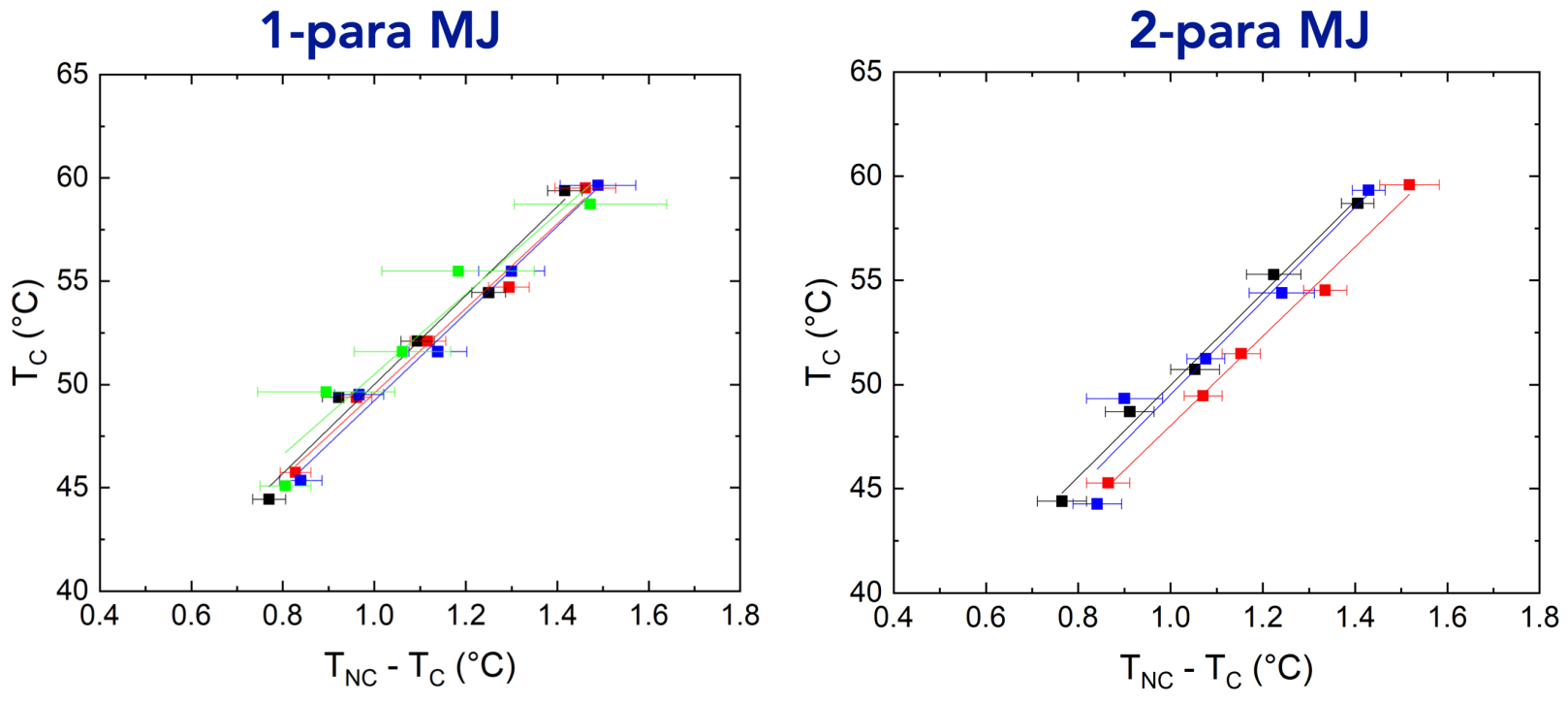

***Figure S20**. $T_C$ vs. $T_{NC}$-$T_C$ plots at 3 or 4 different locations on the SAMs for the **1-para** and **2-para** SAMs. The lines are linear fits to calculate the sample thermal conductance $G_{th}$(SAM/Au).*

| | pristine | | pristine | protonated | deprotanated |
|---|---|---|---|---|---|
| ***1-para*** | 37.2 ± 4.4 | ***1-meta*** | 16.3 ± 4.7 | 24.0 ± 3.5 | 16.4 ± 1.9 |
| ***2-para*** | 39.5 ± 4.6 | ***2-meta*** | 29.0 ± 6.4 | 33.8 ± 6.3 | 10.8 ± 2.6 |

***Table S2**. Mean thermal conductances of the SAMs (in nW/K) of para- and meta-connected molecules (see Fig. 4-b) in the pristine, protonated and deprotonated states.*

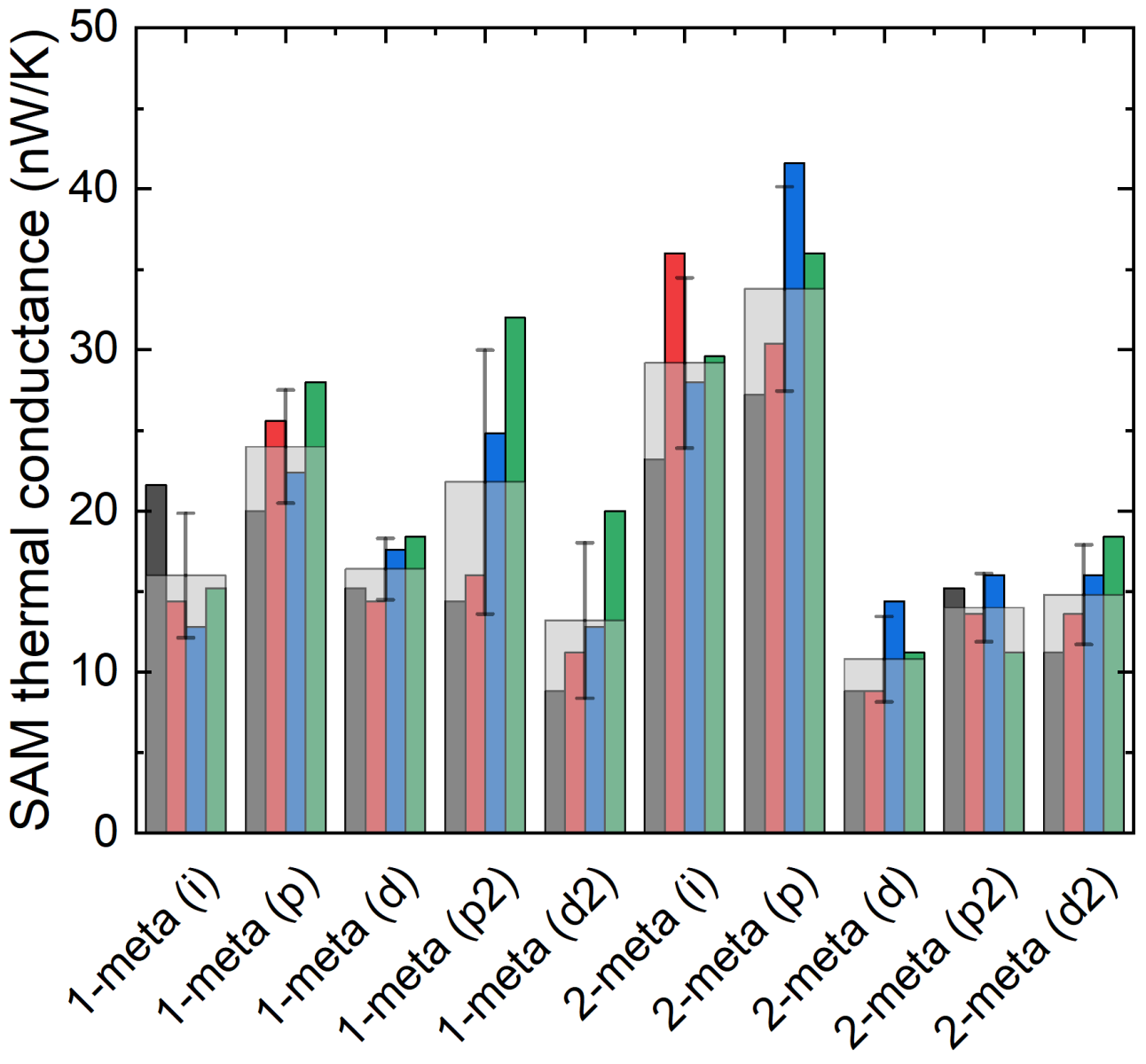

***Figure S21**. Switching of the SAM thermal conductance upon a second protonation/deprotonation cycle (marked as p2 and d2). Data for initial cycle (i,p,d) from Fig. 4-b (main text).*

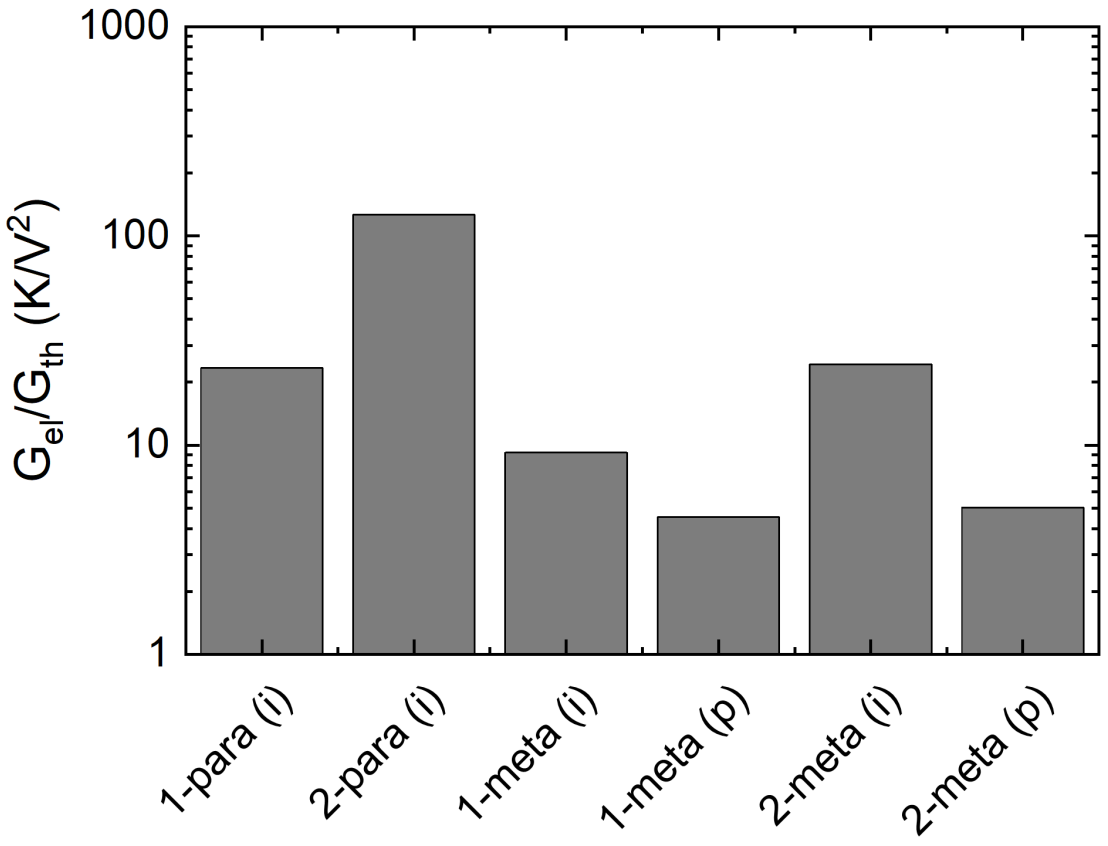

***Figure S22**. Ratios $G_{el}/G_{th}$. The zero-bias electrical conductance $G_{el}$ was calculated from the first derivative of the mean I-V curves of the data sets. The derivatives were slightly smoothed (2nd order polynomial Savitzky-Golay smooth on 20 points) and the conductance taken in the bias window -50/50 mV. The thermal conductances were calculated from the data in Table S2. Since the C-AFM and SThM tip contact areas and the SAM molecule packing densities are not the same in the various SAMs, we have normalized the ratio $G_{el}/G_{th}$ to the single molecule level considering the estimated contact areas (see sections IV.3 and V.2 and the estimated area per molecule in the SAM, see Table S1). The estimated data are shown in Table S3.*

| | $G_{el}$ (nS) | $G_{th}$ (pW/K) | $G_{el}/G_{th}$ (K/V$^2$) |
|---|---|---|---|
| 1-para (i) | 0.17 | 7.4 | 22.9 |
| 2-para (i) | 12.00 | 95.1 | 126.2 |
| 1-meta (i) | 0.47 | 51.6 | 9.1 |
| 1-meta (p) | 0.35 | 77.4 | 4.5 |
| 2-meta (i) | 7.10 | 292.0 | 24.3 |
| 2-meta (p) | 1.70 | 338.0 | 5.0 |

***Table S3.** Equivalent single-molecule electrical and thermal conductances of the MJs, and the corresponding $G_{el}/G_{th}$ ratios reported in Fig. S22.*

# References.